\documentclass[journal,romanappendices,twoside]{IEEEtran}

\pdfoutput=1
\usepackage{cite}
\usepackage{graphicx}
\graphicspath{{fig/}{img/}{fig/pdf/}}
\usepackage{amsfonts}
\usepackage{amssymb,amsmath}
\usepackage{dsfont}
\usepackage{booktabs}
\usepackage{calc}
\usepackage{ifthen}
\usepackage[dvipsnames,usenames]{color}
\usepackage{acronym}
\usepackage{nicefrac}
\usepackage{enumitem}
\usepackage{units}
\usepackage{soul}
\usepackage[absolute]{textpos}
\usepackage{comment}
\usepackage[demo]{pdfpages}

\newcommand{\op}[1]{\mathbm{#1}}

\def\mathbm#1{\text{\fontfamily{cmr}\fontseries{b}\selectfont #1}}

\newcommand{\WF}{\op{W}_\mathrm{f}}
\newcommand{\IWF}{\widetilde{\op{W}}_\mathrm{f}}
\newcommand{\XF}{\op{X}_\mathrm{f}}
\newcommand{\WT}{\op{W}_\mathrm{t}}
\newcommand{\IWT}{\widetilde{\op{W}}_\mathrm{t}}
\newcommand{\XT}{\op{X}_\mathrm{t}}
\newcommand{\WX}{\op{W}_\mathrm{x}}
\newcommand{\IWX}{\widetilde{\op{W}}_\mathrm{x}}
\newcommand{\XX}{\op{X}_\mathrm{x}}
\newcommand{\IXX}{\widehat{\op{X}}_\mathrm{x}}

\newacro{TF}{Time-Frequency}
\newacro{FW}{Frequency Warping}
\newacro{TW}{Time Warping}
\newacro{SWF}{Sample Without Filtering}
\newacro{SAF}{Sample After Filtering}
\newacro{NUFFT}{Non Uniform Fast Fourier Transform}
\newacro{CQT}{Constant-Q Transform}

\newcommand{\supmat}{true}
\newcommand{\addnum}{false}
\newcommand{\addtab}{false}
\newcommand{\addpar}{false}
\newcommand{\addapp}{false}

\newcommand{\appref}[1]{\ifthenelse{\boolean{\addapp}}{#1}{}}
\newcommand{\nonnum}[1]{\ifthenelse{\boolean{\addnum}}{}{#1}}

\begin{document}

\title{Time Warping and Interpolation Operators\\for Piecewise Smooth Maps}

\author{Salvatore~Caporale and Yvan~Petillot
\thanks{S. Caporale  and Y. Petillot are with the Institute of Sensors, Signals and Systems, Heriot-Watt University, Edinburgh, Scoltland, UK (e-mail: s.caporale@hw.ac.uk; y.r.petillot@hw.ac.uk).}}


\markboth%
{Caporale \MakeLowercase{\textit{et al.}}: Time Warping and Interpolation Operators for Piecewise Smooth Maps}
{Caporale \MakeLowercase{\textit{et al.}}: Time Warping and Interpolation Operators for Piecewise Smooth Maps}

%


\maketitle

\begin{abstract}
A warping operator consists of an invertible axis deformation applied either in the signal domain or in the corresponding Fourier domain. Additionally, a warping transformation is usually required to preserve the signal energy, thus preserving orthogonality and being invertible by its adjoint. Initially, the design of such operators has been motivated by the idea of suitably generalizing the properties of orthogonal time-frequency decompositions such as wavelets and filter banks, hence the energy preservation property was essential. Recently, warping operators have been employed for frequency dispersion compensation in the Fourier domain or the identification of waveforms similarity in the time domain. For such applications, the energy preservation requirement can be given up, thus making warping a special case of interpolation. In this context, the purpose of this work is to provide analytical models and efficient computational algorithms for time warping with respect to piecewise smooth warping maps by transposing and extending a theoretical framework which has been previously introduced for frequency warping. Moreover, the same approach is generalized to the case of warping without energy preservation, thus obtaining a fast interpolation operator with analytically defined and fast inverse operator.
\end{abstract}


\IEEEpeerreviewmaketitle

\section{Introduction}
\label{sec:intro}


\IEEEPARstart{S}{ignal} processing spans a large variety of theoretical and application frameworks. Time-frequency tools address the necessity of representing signals in suitable domains in order to highlight specific features or properties. The design and introduction of these tools can be driven by novel abstract concepts and paradigms, emerging applications or the sake of computability and implementability.  Warping techniques have been introduced as a new theoretical concept pursuing the idea of generating new \ac{TF} unitary representations, starting from already known ones such as wavelet analysis or filter banks~\cite{Oppenheim1972,Baraniuk1995}. In this framework, rather than designing \ac{TF} analysis tools specifically suited to a class of signals or aiming to extract specific \ac{TF} features, an orthogonal deformation is applied to either the time or the frequency axis of the input signal followed by a standard \ac{TF} analysis tool. This operation is equivalent to composing the warping operator with the \ac{TF} operator. Since the composition of two subsequent orthogonal operators is still an orthogonal operator, a new orthogonal operator featuring the prescribed properties is obtained. This idea has been partially exploited but has not revealed any disruptive innovation~\cite{Makur2001,Makur2008,Twaroch1998}. Although this approach is \emph{conceptually} appealing, many limitations arises from both \emph{application} and \emph{computational} point of view. In more detail, when dealing with \ac{FW}, operations over the frequency axis implies the knowledge of the entire signal, hence working on finite length or windowed signals is mandatory. When dealing with finite spaces, warping operators can be designed only as \emph{tight frames} rather than unitary operators~\cite{Yeh1997}, meaning that the obtained representation features some redundancy. Finally, with regard to computation, some issues arise when moving from the definition of warping in continuous domains to its implementation in discrete domains.

Despite the above considerations, warping has been effectively applied in various signal processing areas. For instance, current successful and popular implementations of warping attempt to compensate for inherent deformation occurring on either the time or the frequency axis because of physical phenomena such as dispersive propagation~\cite{Papandreou1993,Hlawatsch1997,DeMarchi2009,DeMarchi2009b,Papandreou-Suppappola2001,Caporale2012,Caporale2015a}. In those applications, theoretical and computational aspects become secondary as the knowledge of the physical parameters or functions, upon which the design of the warped operator is based, is often not accurate enough to require properties such as orthogonality and perfect reconstruction~\cite{Bonnel2013,Bonnel2016}. Moreover, as the main goal is to invert an axis deformation, the energy preservation requirement does not always apply. Other applications include the Constant-Q Transform, where the invertibility is a major issue~\cite{holighaus2013} and image compression~\cite{asghari2014}.

Previous works by the authors about \ac{FW}~\cite{Caporale2007,Caporale2008,Caporale2010b} mainly aimed at solving implementation issues. In fact, a thorough investigation of computational aspects and signal representation issues involved in the design of \ac{FW} as a \emph{frame} has been pursued. More specifically, with the aim of obtaining an approximation of \ac{FW} being more accurate than the one obtained by simply sampling the frequency axis, an interesting theoretical framework about the representation of signals having non-smooth Fourier transforms has been identified together with a fast algorithm for \ac{FW} computation. In addition, the same framework has been exploited for obtaining an analytical representation of the dual frame allowing perfect and fast reconstruction. 

The work we here propose aims to move to \ac{TW} the mathematical framework which has been identified for \ac{FW}, i.e. direct and inverse transform with respect to piecewise smooth warping maps. In this context, moving from the Fourier domain to the time domain arises theoretical and computational differences about the continuous representation of the warped axis and about design constraints which will be dealt with in the paper. 
As a natural completion of this framework, we also extend the same theory to the identify a time interpolation operator and its inverse.

The paper is organized as follows. Section~\ref{sec:goals} reviews basic concepts about warping operators and describes the problems being tackled in this work. In Section~\ref{sec:freq2time} the transition from \ac{FW} to \ac{TW} is detailed whereas in Section~\ref{sec:warp2interp} warping is revisited as interpolation. Finally, some experimental results and conluding remarks are shown in Section~\ref{sec:results} and Section~\ref{sec:concl} respectively.

\section{Warping Review and Goals}
\label{sec:goals}

First a review of the basics of warping are presented~\cite{Caporale2009a,Caporale2014}, introducing a new perspective in order to focus on the difference which arises when moving from \ac{FW} to \ac{TW}. Moreover we also introduce the mathematical model for employing the warping framework to perform interpolation and its inverse.

\subsection{Warping and interpolation in continuous spaces}
\label{subsec:contspace}

The core transformation of warping is described by the following deformation operator applied to either the time or the frequency axis
\begin{equation}\label{eq:warpker}
	\mathfrak{W}(x,y)=(Dw(x))^{\nicefrac{1}{2}}\,\delta(w(x)-y)\qquad{}x,y\in\mathds{R}
\end{equation}
where $w$ is a bijection, hence its derivative $D{w}$ is always finite and positive, thus the inverse map $w^{-1}$ also exists. Operator $\mathfrak{W}$ acts as \emph{warping} if the integration is performed with respect to the second variable of the kernel
\begin{eqnarray}\label{eq:warpsig}
	[\mathfrak{W}s](x)
	&=&
	\int_{\mathds{R}}(Dw(x))^{\nicefrac{1}{2}}\,\delta(w(x)-y)\,s(y)\,dy
	\nonumber
	\\
	&=&
	(Dw(x))^{\nicefrac{1}{2}}\,s(w(x))
\end{eqnarray}
whereas it acts as \emph{unwarping} if applied to the first variable, i.e. when the adjoint operator is considered (operator \eqref{eq:warpker} is real so the transpose is equal to the adjoint)
\begin{eqnarray}\label{eq:iwarpsig}
	[\mathfrak{W}^\dag{}s](y)&=&\int_{\mathds{R}}(Dw(x))^{\nicefrac{1}{2}}\,\,\delta(w(x)-y)\,s(x)\,dx
	\nonumber
	\\
	&=&\int_{\mathds{R}}(Dw^{-1}(x))^{\nicefrac{1}{2}}\,\delta(z-y)s(w^{-1}(z))\,dz
	\nonumber
	\\[5pt]
	&=&
	(Dw^{-1}(x))^{\nicefrac{1}{2}}\,s(w^{-1}(y))
\end{eqnarray}
hence the direct operator built from the inverse map is equal to the adjoint (or inverse) operator built from the direct map. By setting $v=w^{-1}$ and $\mathfrak{V}$ equal to the operator corresponding to $v$, we have $\mathfrak{V}=\mathfrak{W}^\dag$. This property becomes more intuitive if one gives up to the orthogonalization factor and considers the neat interpolation problem. To show this, let us introduce the generalized operator
\begin{equation}\label{eq:warpkernonort}
	\mathfrak{W}^{(b)}(x,y)=(Dw(x))^{b}\,\delta(w(x)-y)\qquad{}x,y\in\mathds{R}
\end{equation}
with $b$ representing a generic power $b\in[0,1]$, such that the simple interpolation can be referred as $\mathfrak{W}^{(0)}$. The natural way for inverting $\mathfrak{W}^{(0)}$ would be to apply $\mathfrak{V}^{(0)}$. Less intuitively, $\mathfrak{W}^{(0)}$ can be equivalently inverted by applying $\mathfrak{W}^{(1)\dag}$, i.e.
\begin{equation}
	\mathfrak{W}^{(1-b)\dag}\mathfrak{W}^{(b)}=\mathfrak{I}
	\nonumber
\end{equation}
being $\mathfrak{I}$ the identity operator. In fact, the composition can be computed by
\begin{multline}
	[\mathfrak{W}^{(1-b)\dag}\mathfrak{W}^{(b)}](x,y)=
	\\
	\int_{\mathds{R}}(Dw(z))^{1-b}\,\delta(w(z)-x)\,(Dw(z))^{b}\,\delta(w(z)-y)\,dz=
	\\
	\int_{\mathds{R}}\delta(u-x)\,\delta(u-y)\,du
	=\delta(x-y).
	\label{eq:warpnonortid}
\end{multline}
Obviously for these continuous operators it must hold $\mathfrak{W}^{(1)\dag}=\mathfrak{V}^{(0)}$. Instead, when coping with discrete-time signals, considering the approximated operator derived from $\mathfrak{W}^{(1)\dag}$ rather than from $\mathfrak{V}^{(0)}$ brings to different results and accuracies. Section~\ref{sec:warp2interp} addresses the  problem of approaching the inverse of the interpolation operator.

\subsection{Warping in periodic spaces}
\label{subsec:perspace}

The continuous approach described in \ref{subsec:contspace} has two major limitations. The warping operator \eqref{eq:warpker} and its orthogonality rely on an infinite continuous axis. As a first step, one has to switch from an infinite axis to a limited interval in order to deal with the fact that input signals are known on a limited time interval for \ac{TW} or are band-limited signals for \ac{FW}. Relaxing the continuity is more delicate and will be dealt with in \ref{subsec:perspace}.

As far as \ac{FW} is concerned, working with a limited axis is quite natural. In fact, a time-continuous band-limited signal can be safely sampled its spectrum will be continuous and periodic. In addition, it is required that the number of non-zero samples is finite, so that the signal is actually manageable. So, the \ac{FW} operator designed for discrete-time signals always corresponds to a periodic-wise map. The periodic-wise map, with respect to a normalized period of length $1$ verifies the following property
\begin{equation}\label{eq:periodmap}
	w(x+k)=w(x)+k
	\qquad
	k\in\mathds{Z}.
\end{equation}
For \ac{FW}, the warping function must be also odd in order to transform real signals into real signals~\cite{Caporale2009a}. The key concept introduced about \ac{FW} is that , although the map is designed on a single period, its properties have to be evaluated on its periodic-wise extension. So, a map being smooth on a single period becomes only piecewise smooth when considered on the entire frequency axis. 

\begin{figure}[t]
	\centering%
	\includegraphics[scale=.75,trim=0 10pt 0 10pt]{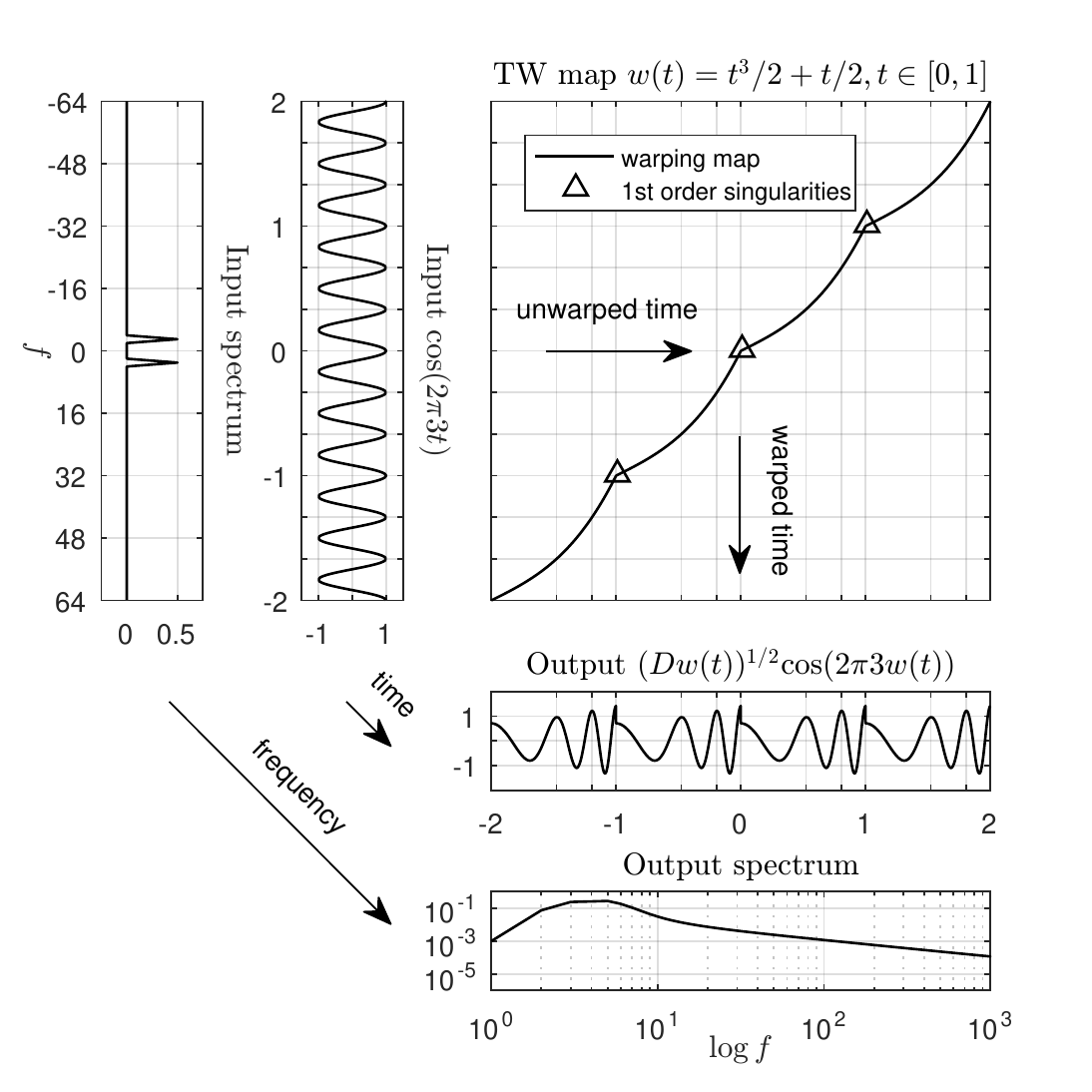}%
	\caption{%
		Piecewise smooth warping map applied to a sinusoidal input in the continuous periodic time domain. Fixed points of the warping map are placed in $t=k,k\in\mathds{Z}$. Time warping maps do not have any constraints about junction points between contiguous intervals, so the map is only $\mathcal{C}^0$ and output is not continuous because of the orthogonalizing factor $(Dw)^{\nicefrac{1}{2}}$. The input line spectrum is transformed into a $\nicefrac{1}{f}$ decaying spectrum.
	}%
	\label{fig:warptime}
\end{figure}

\begin{figure}[t]
	\centering%
	\includegraphics[scale=.75,trim=0 10pt 0 10pt]{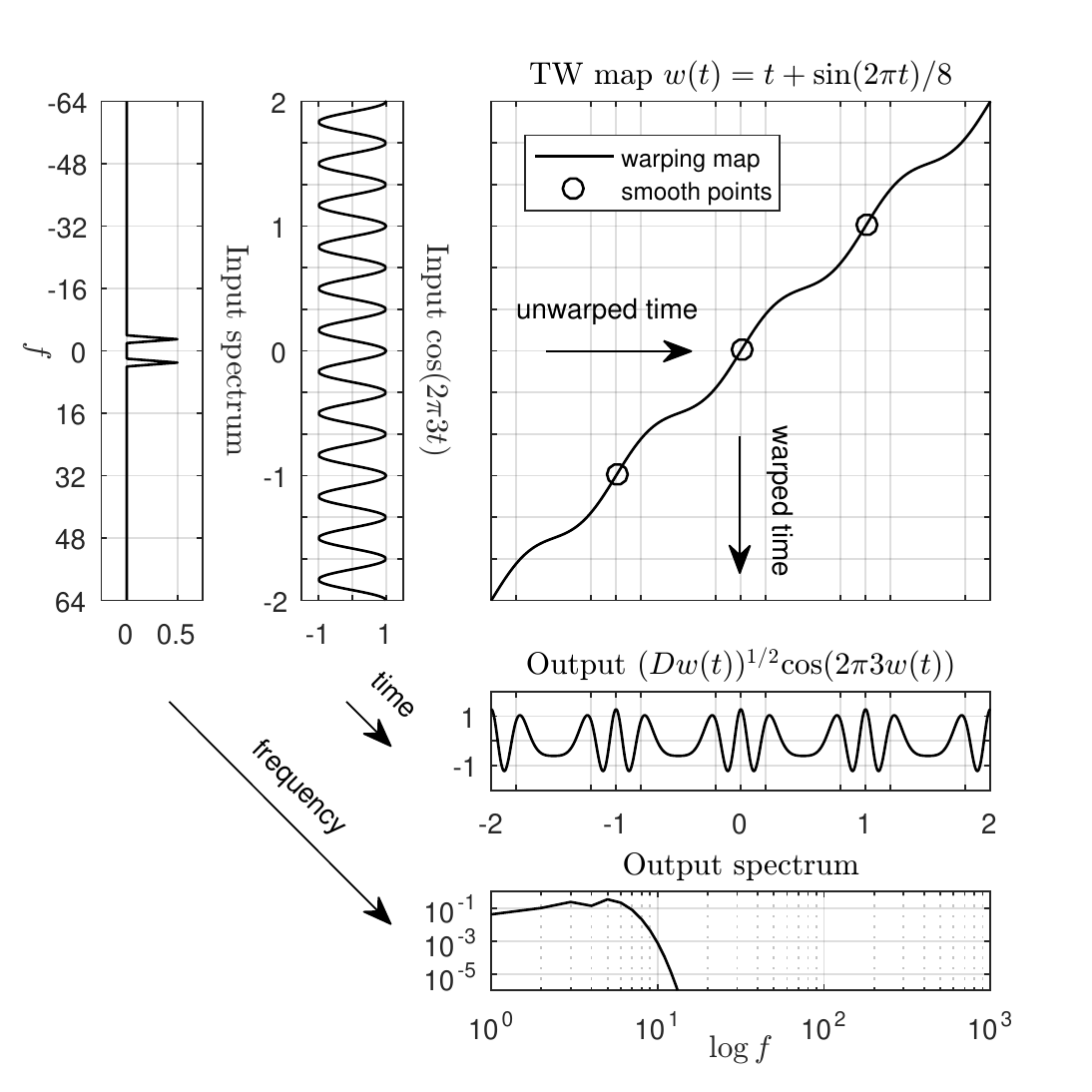}%
	\caption{%
		Smooth warping map applied to a sinusoidal input in the continuous periodic time domain. Fixed points of the warping map are placed in $t=k,k\in\mathds{Z}$. Junction points between periods are smooth, hence the output is also globally smooth. The resulting output spectrum features an exponential decay, hence the output can be considered band-limited and can be correctly represented by means of a proper sampling.
	}%
	\label{fig:warpsmooth}
\end{figure}

With reference to \ac{TW}, let us now consider a discrete-time signal in time domain. A continuous representation could be obtained by ideal interpolation, i.e. by applying an ideal rectangular filter to its spectrum. A limited time interval could be isolated by time domain windowing. This approach would be formally correct but the resulting time-continuous representation would depend on samples not belonging to the considered time-window. In order to be able to represent \ac{TW} as an algebraic operation with respect to an input signal of finite dimension, we consider as continuous-time representation the circulant interpolation. By doing so, as for \ac{FW}, \ac{TW} can be described as a deformation of the entire time axis performed by means of a suitable periodic-wise map. The time axis deformation is represented in Fig.~\ref{fig:warptime} with respect to a single time domain sinusoidal component. A time map being smooth with respect to a single period is guaranteed to be only $\mathcal{C}^0$ (continuous with non-continuous derivatives) with respect to the entire time axis. Hence, the orthogonalizing factor $(Dw)^{\nicefrac{1}{2}}$ in equation \eqref{eq:warpker} and so the warped signal cannot be guaranteed to be $\mathcal{C}^0$ unless the warping map features special regularities conditions at the period boundaries. Hence, \ac{TW} maps are generally assumed to be \emph{piecewise smooth}. As an example, in Fig.~\ref{fig:warpsmooth}, the process of warping the same signal as in Fig.~\ref{fig:warptime} by means of a globally smooth warping map is represented. The resulting spectrum has a fast decay, thus the signal is practically bandlimited and the formal description in the periodic space does not involve any advantage. As previously done for \ac{FW} in~\cite{Caporale2009a,Caporale2014}, the aim of this paper is to work with piecewise smooth maps like splines as they allow for a much flexible design.

\subsection{Warping in finite-dimensional spaces}
\label{subsec:finitespace}

The operations described so far applies to finite-support domains but still involve continuous integration as in equations \eqref{eq:warpsig} and \eqref{eq:iwarpsig}, which obviously is not computable with a finite procedure or representable with a finite series unless the continuous function $(Dw(x))^{\nicefrac{1}{2}}s(w(x))$ is isomorphic with a finite space. The intuitive way to decrease the space dimension from infinite to finite is to perform a sufficiently dense sampling on $(Dw(x))^{\nicefrac{1}{2}}s(w(x))$. This approach will be referred to as \ac{SWF}. If the warping map is only piecewise smooth over the whole axis, the function $(Dw(x))^{\nicefrac{1}{2}}\,s(w(x))$ spectrum has a polynomial decay because of the warping function singularities, as represented in Fig.~\ref{fig:warptime}. Hence, from a pure theoretical point of view $(Dw(x))^{\nicefrac{1}{2}}\,s(w(x))$ is not band-limited and cannot be sampled without aliasing. The less intuitive but more accurate way to perform the space dimension reduction is to first apply a filter over $(Dw(x))^{\nicefrac{1}{2}}\,s(w(x))$ followed by a suitable dense sampling. This strategy will be referred to as \ac{SAF}. The \ac{SWF} can be simply implemented by means of \ac{NUFFT} algorithms~\cite{dutt:1368,Liu1998,Liu1998a,Fessler2003}, whereas the \ac{SAF} requires some further processing~\cite{Caporale2009a}.

For \ac{FW}, the function $(Dw(x))^{\nicefrac{1}{2}}\,s(w(x))$ is complex and its spectrum is actually a time domain interval. Conversely, for \ac{TW} $(Dw(x))^{\nicefrac{1}{2}}\,s(w(x))$ is in time domain, hence its spectrum is actually a proper frequency spectrum. In \ref{subsec:freq2timedom} we will detail afterwards how to select the filtering \emph{band} for \ac{TW} and \ac{FW}.

According to what we have described so far, \ac{TW} can be modelled in the following way. First we introduce supporting Fourier and windowing/filtering operators. As a convention, all operators applied to discrete-time domains will be represented with vector notation, whereas warping operators applied to continuous-time domains have been represented by gothic letters. We first introduce the Fourier series
\begin{equation}\label{eq:fourierseries}
	\op{F}(k,x)=\mathrm{e}^{-j2\pi{}kx}
	\quad
	x\in[0,1),\,k\in\mathds{Z}
\end{equation}
and the Discrete Fourier Transform operator
\begin{equation}\label{eq:fouriertransf}
	\op{F}_N(k,n)=N^{-\nicefrac{1}{2}}\mathrm{e}^{-j2\pi{}kn/N}
	\quad
	n,k\in\mathds{Z}_N
\end{equation}
where by now $\mathds{Z}_N$ is a suitably defined set of $N$ contiguous integer (details will be given in \ref{subsec:freq2timedom}). We also introduce the warped Fourier series
\begin{equation}\label{eq:fourierwarp}
	\op{F}_w(k,x)=(Dw(x))^{\nicefrac{1}{2}}\,\mathrm{e}^{-j2\pi{}kw(x)}
	\quad
	x \in[0,1),\,k\in\mathds{Z}_N
\end{equation}
and the time windowing/filtering operator $\op{L}_M$
\begin{equation}\label{eq:window}
	\op{L}_M(l,k)=\delta(l-k)
	\quad
	l\in\mathds{Z}_M,k\in\mathds{Z}.
\end{equation}
Finally, the approximate \ac{TW} operator can be computed by performing the warped circulant interpolation $\op{F}_w^{\dag}\op{F}_N$, evaluating its Fourier series by $\op{F}$, selecting $M$ spectrum component, $M>N$, and going back to the time domain by $\op{F}_M^\dag$, that is
\begin{equation}\label{eq:warptime}
	\op{W}_\mathrm{t}=\op{F}_M^\dagger\op{L}_M\op{F}\,\op{F}_w^{\dag}\op{F}_N
\end{equation}
where the $\mathrm{t}$ subscript stays for time.

As far as \ac{FW} is concerned, the problem is slightly simpler as there is no need to perform a circulant interpolation and to go back to the warped domain. We have
\begin{equation}\label{eq:warpfreq}
	\op{W}_\mathrm{f}=\op{L}_M\op{F}^*\op{F}_w^{\prime}
\end{equation}
where $\op{F}_w^{\prime}$ represents a warped Fourier transform for a discrete-time signal and $\op{F}^*$ the inverse Fourier transform.

\subsection{Paper goals and results}
\label{subsec:goal}

For \ac{FW}, a way to approach \ac{SAF} by compensating aliasing on \ac{SWF} has been modelled and analysed in~\cite{Caporale2009a}. A decomposition allowing for a fast computation has been also provided. In~\cite{Caporale2014}, the dual operator $\widetilde{\op{W}}_\mathrm{f}^{}$ such that $\widetilde{\op{W}}_\mathrm{f}^{\dag}\op{W}_\mathrm{f}^{}=\op{I}$ has been identified by a analytical model. The main goal of this work is to transpose the results obtained for \ac{FW} into \ac{TW} together with some generalizations and expansions. In more detail, in Section \ref{sec:freq2time} we will focus on (i) how to apply to $\op{W}_\mathrm{t}$ the model and algorithm obtained for $\op{W}_\mathrm{f}$, (ii) how to transpose the algorithm for $\widetilde{\op{W}}_\mathrm{f}$  into $\widetilde{\op{W}}_\mathrm{t}$ and (iii) how the choice of input and output domains differently impact on \ac{TW} and \ac{FW}.
With respect to (iii), we will also extend \ac{FW} to input and output domains which have not been covered in previous works. Furthermore, in Section \ref{sec:warp2interp}, according to the notation used in equation \eqref{eq:warpkernonort}, we will also detail how to generalize operator $\op{W}_\mathrm{t}$, which could be also referred to as $\op{W}_\mathrm{t}^{(\nicefrac{1}{2})}$, into the generic operator $\op{W}_\mathrm{t}^{(b)}$, with special reference to the pure interpolating operator $\op{W}_\mathrm{t}^{(0)}$ and its inverse $\widetilde{\op{W}}_\mathrm{t}^{(1)}$.

\section{From Frequency Warping to Time Warping}
\label{sec:freq2time}

Rather than first providing a review of \ac{FW} and then redefining the model for \ac{TW}, we provide results obtained for \ac{FW} together with their modifications for \ac{TW} and generalizations.

\subsection{Model transition for the direct operator}
\label{subsec:freq2timedir}

The core idea of the method which have been proposed in~\cite{Caporale2009a} is to obtain the operator $\WF$ corresponding to \ac{SAF} approach by correcting the operator obtained by \ac{SWF}. To describe both \ac{TW} and \ac{FW} with the same approach, we introduce the following $\infty\times\infty$ shift-variant operator
\begin{multline}\label{eq:warpinf}
	\op{W}^{}(m,n)=\int_0^1(Dw(x))^{\nicefrac{1}{2}}\,e^{j2\pi(mf-nw(x))}dx
	\\
	n,m\in\mathds{Z}
\end{multline}
which features a \emph{center} of symmetry for $n=0$ and $m=0$ as it can be easily verified that $\op{W}(-m,-n)=\op{W}^*(m,n)$. Operator \eqref{eq:warpinf} is a \emph{unitary} operator, i.e. it satisfies the requirements for getting the exact inversion by means of its adjoint as it represents a pure transposition of \eqref{eq:warpker} into discrete-time domains according to what has been described in \ref{subsec:perspace}. According to \eqref{eq:warpfreq}, $\op{W}_\mathrm{f}$ can be obtained by \emph{windowing} both its input and its output. By noticing $\op{F}^*\op{F}_w'=\op{W}\,\op{L}_N'$, we can write
\begin{equation}\label{eq:warpfreqwind}
	\op{W}_\mathrm{f}=\op{L}_M\op{W}\,\op{L}_N^\prime
	\nonumber
\end{equation}
where, operator $\op{L}_M$ selects the a specific set $\mathds{Z}_{N,L_N}$ rather than just the generic $\mathds{Z}_{N}$ which have been used in relation to equations \eqref{eq:fouriertransf}-\eqref{eq:window}. Hence, $\mathds{Z}_{N,L_N}$ is
\begin{equation}\label{eq:inputshift}
	\mathds{Z}_{N,L_N}=\{-L_N,\ldots,N-L_N-1\}.
\end{equation}
and $L_N$ is usually taken in the interval $[0,N-1]$ in order to include the center of symmetry. In previous work on \ac{FW} only the symmetrical case $\mathds{Z}_{N,N/2}$ with $N$ even has been considered. The extension to a generic $L_N$ is dealt with in \ref{subsec:freq2timedom}. In order to relate $\op{W}$ to $\op{W}_\mathrm{t}$, we highlight that $\op{F}\op{F}_w^{\dag}=(\op{F}^*\op{F}_w')^*$, then equation \eqref{eq:warptime} can be rewritten as
\begin{equation}\label{eq:warptimewind}
	\op{W}_\mathrm{t}=\op{F}_M^\dagger\op{L}_M\op{W}^*\op{L}_N^\prime\op{F}_N.
\end{equation}
Hence, we point out the following principle. Every result obtained for \ac{FW} involving $\op{W}$ can be employed for \ac{TW} by simply  applying a conjugation to $\op{W}$ and including it between discrete Fourier transform and its inverse of size $N$ and $M$ respectively.

The \ac{SWF} approximation for \ac{FW} is obtained by sampling the integral in \eqref{eq:warpinf} in $M$ points and will be referred to as $\op{X}_\mathrm{f}$
\begin{multline}\label{eq:warpfreqaliaslong}
	\op{X}_\mathrm{f}(m,n)=
	\frac{1}{M}\sum_{k=0}^{M-1}\,(Dw(k/M))^{\nicefrac{1}{2}}\,e^{j2\pi(mk/M-nw(k/M))}
	\\
	m\in\mathds{Z}_{M,L_M},n\in\mathds{Z}_{N,L_N}
\end{multline}
and can be described by
\begin{equation}\label{eq:warpfreqalias}
	\op{X}_\mathrm{f}=\op{F}_M^*\,\op{F}_{w,M}^{\prime}
\end{equation}
where $\op{F}_{w,M}$ is the Warped Discrete Fourier Transform
\begin{equation}\label{eq:fourierwarptransf}
	\op{F}^{}_{w,M}(k,m)=M^{-\nicefrac{1}{2}}(Dw(m/M))^{\nicefrac{1}{2}}\,\mathrm{e}^{-j2\pi{}kw(m/M)}
	\nonumber
\end{equation}
which can be computed by means of \ac{NUFFT} algorithm~\cite{dutt:1368,Liu1998,Liu1998a,Fessler2003}. For $\op{X}_\mathrm{t}$, by applying the principle identified in \eqref{eq:warptimewind}, we have
\begin{eqnarray}\label{eq:warptimealias}
	\op{X}_\mathrm{t}
	&=&
	\op{F}_M^\dagger\op{L}_M\op{X}_\mathrm{f}^*\op{L}_N\op{F}_N
	\nonumber
	\\
	&=&
	\op{F}_M^\dagger(\op{F}_M^*\,\op{F}_{w,M}^\prime)^*\op{F}_N
	\nonumber
	\\
	&=&
	\op{F}_{w,M}^\dag\op{F}_N
\end{eqnarray}
where $\op{L}_N$ and $\op{L}_M$ have been neglected as both input and output of $\op{X}_\mathrm{f}$ are already band-limited. The above result can be considered obvious, as it represents the obvious way to perform a warped circulant interpolation. Operator $\op{X}_\mathrm{t}$ and $\op{X}_\mathrm{f}$, as well as $\op{W}_\mathrm{t}$ and $\op{W}_\mathrm{f}$, are computed in a similar way but have quite different characteristic as $\op{X}_\mathrm{t}$ and $\op{W}_\mathrm{t}$ behave as circulant warped interpolators whereas $\op{X}_\mathrm{f}$ and $\op{W}_\mathrm{f}$ appear as dispersive delaying operators.

Following the equivalence which has been highlighted in \eqref{eq:iwarpsig}, the sampled \ac{TW} operator corresponding to the inverse map $v=w^{-1}$ is also introduced
\begin{equation}\label{eq:warptimealiasinv}
	\widehat{\op{X}}_\mathrm{t}=
	(\op{F}_{v,N}^\dag\op{F}_M)'=\op{F}_M^\prime\op{F}_{v,N}^*.
\end{equation}
As it has been anticipated in Section \ref{sec:goals}, although for the operators in continuous spaces corresponding to $w$ and $v$ it holds $\mathfrak{V}=\mathfrak{W}^\dag$, for the sampled operators we have $\op{X}_\mathrm{t}\neq\widehat{\op{X}}_\mathrm{t}$. Nevertheless, $\widehat{\op{X}}^\prime_\mathrm{t}$ can be still used as an approximation of the inverse operator, i.e. $\widehat{\op{X}}^\prime_\mathrm{t}\op{X}_\mathrm{t}\simeq\op{I}$. An accuracy comparison will be provided later in this Section. 

\begin{figure}[t]
	\centering%
	\includegraphics[scale=.75,trim=0 10pt 0 10pt]{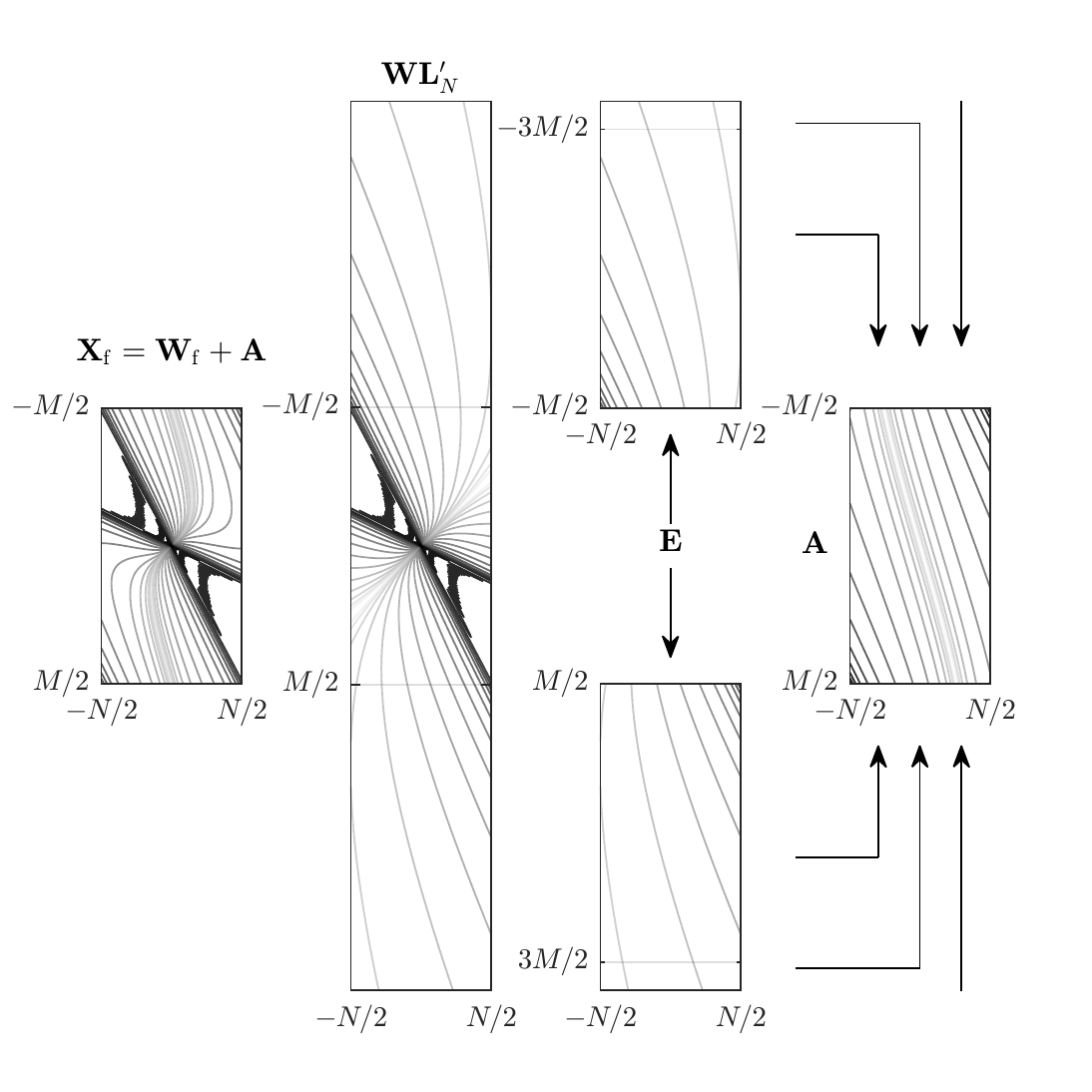}%
	\caption{%
		Schematic representation of the computational strategy for obtaining matrix $\op{W}_\mathrm{f}$ \eqref{eq:warpfreqwind} corresponding to the \ac{SAF} approach by correcting matrix $\op{X}_\mathrm{f}$ \eqref{eq:warpfreqalias} corresponding to the \ac{SWF} approach. Logarithmic absolute values of the matrix items are represented by isolines. On the left, the typical structure of matrix $\op{W}$ can be observed: most significant items are enclosed between two lines whose slope are $\mathrm{min}Dw$ and $\mathrm{max}Dw$ respectively. Matrix $\op{E}$ has low rank as a consequence of the presence of singularities in the warping map. Matrix $\op{A}$, being obtained by periodic summation over $\op{E}$ columns, has also low rank. For representation purposes, both $\op{W}\op{L}_{N}^\prime$ and $\op{E}$ columns, having infinite length, have been truncated.
	}%
	\label{fig:warpalg}
\end{figure}

The computational strategy for $\op{W}_\mathrm{f}$ consists in first finding a model for the decaying tails of $\op{W}$ for $m\to\pm\infty$, then using this model to compute aliasing to correct operator $\op{X}_\mathrm{f}$. The error operator which is introduced when switching from operator $\op{W}\op{L}^\prime_N$ to operator $\WF$ is represented by
\begin{equation}\label{eq:tails}
	\op{E}(m,n)=
	\left\{%
	\begin{array}{ll}
	0 & \quad m\notin\mathds{Z}_{M,L_M}\\
	\op{W}\op{L}^\prime_N(m,n) & \quad m\in\mathds{Z}_{M,L_M}
	\end{array}%
	\right.
\end{equation}
and aliasing is obtained by periodic sum over $\op{E}$
\begin{equation}\label{eq:alias}
	\op{A}(m,n)=\sum_{k\in\mathds{Z}}\op{E}(m-kM,n)
\end{equation}
such that $\WF=\XF-\op{A}$. This decomposition is schematically represented in Fig.~\ref{fig:warpalg}, where also the typical sparsity pattern and decay of matrix $\op{W}$ is highlighted. We refer to the singularities of the considered piecewise smooth map as $\xi_i, i=1,\ldots,I$. Under certain conditions which will be detailed in \ref{subsec:freq2timedom}, $\op{E}$ and $\op{A}$ can be factorized as follows
\begin{alignat}{2}
	\op{E}
	&=\sum_{i=1}^I\op{E}_i
	&&=
	\sum_{i=1}^I\op{P}_i\op{Y}\,\op{S}_i\op{V}\op{Q}_i
	\label{eq:tailsdec}
	\\
	\op{A}
	&=
	\sum_{i=1}^I\op{A}_i
	&&=
	\sum_{i=1}^I\op{P}_i\op{U}\,\op{S}_i\op{V}\op{Q}_i
	\label{eq:aliasdec}
\end{alignat}
where $\op{P}_i$ and $\op{Q}_i$ are diagonal matrices having as main diagonal $\mathrm{e}^{j2\pi m\xi_i}$ and $\mathrm{e}^{-j2\pi nw(\xi_i)}$  respectively, $\op{S}_i$ is a lower triangular kernel matrix depending on the various order derivatives of the warping map on $\xi_i$ (see equations \eqref{eq:kernelS}-\eqref{eq:kernelJ}) while $\op{V}$, $\op{Y}$ and $\op{U}$ are real matrixes obtained by sampling fixed functions not depending on the warping map. $\op{V}$, $\op{Y}$ and $\op{U}$ will be reviewed in \ref{subsec:freq2timedom} according to the general input and output indexing \eqref{eq:inputshift} introduced in this work (see equations \eqref{eq:npow}, \eqref{eq:mpow} and \eqref{eq:mcot}).

Hence, by following the same principle as in \eqref{eq:warptimewind} we have
\begin{equation}\label{eq:warptimefromalias}
	\op{W}_\mathrm{t}=\op{X}_\mathrm{t}-\op{F}_M^{\dag}\op{A}^*\op{F}_N.
\end{equation}
By taking advantage of \eqref{eq:aliasdec}, the generic time aliasing component referred to $\op{A}_i$ can be rewritten as
\begin{eqnarray}
	\op{F}_M^{\dag}\op{A}_i^*\op{F}_N\!\!
	&=&\!\!
	\op{F}_M^{\dag}\op{P}_i^*\op{U}\,\op{S}_i^*\op{V}\,\op{Q}_i^*\op{F}_N
	\nonumber
	\\
	&=&\!\!
	(\op{F}_M^{\dag}\op{P}_i^*\op{F}_M)(\op{F}_M^{\dag}\op{U})\op{S}_i^*(\op{V}\op{F}_N)(\op{F}_N^\dag\op{Q}_i^*\op{F}_N)
	\label{eq:timealias}
\end{eqnarray}
so that $\op{U}$ and $\op{V}$, being constant matrixes which may be precomputed, can be replaced with their Fourier transforms performed along columns and rows respectively, whereas $\op{P}_i$ and $\op{Q}_i$, being modulations, are turned into circular shifts.

\subsection{Model transition for the dual operator}
\label{subsec:freq2timeinv}

Operator $\WF$ is designed in such a way that $\WF^\dag$ provides a very good approximation of the dual operator. Nevertheless, the exact dual operator of $\op{W}_\mathrm{f}$ has been found in~\cite{Caporale2014} by exploiting the Neumann series
\begin{equation}\label{eq:warpfreqdual}
	\widetilde{\op{W}}_\mathrm{f}=\op{W}_\mathrm{f}\sum_{k=0}^{\infty}(\op{I}-\op{W}_\mathrm{f}^\dag\op{W}_\mathrm{f})^k
\end{equation}
and recalling that $\op{I}-\op{W}_\mathrm{f}^\dag\op{W}_\mathrm{f}=\op{E}^\dagger\op{E}$. By further calculations one gets
\begin{equation}\label{eq:warpfreqdualdec}
	\widetilde{\op{W}}_\mathrm{f}=
	\op{W}_\mathrm{f}
	(\op{I}+\op{H}^\dagger\op{Z\op{H}})
\end{equation}
where $\op{H}$ is a block column matrix whose items are $\op{S}_i\op{V}\,\op{Q}_i$ and $\op{Z}$ is deterministically obtained by $\op{H}$, $\op{Y}$ and $\op{P}_i$. In a similar fashion, we have
\begin{equation}\label{eq:warptimedualdec}
	\widetilde{\op{W}}_\mathrm{t}=
	\op{W}_\mathrm{t}
	(\op{I}+\op{F}_N^\dag\op{H}^\prime\op{Z}^*\op{H}^*\op{F}_N)
\end{equation}
which can be obtained again by applying the principle highlighted in \eqref{eq:warptimewind}
\begin{eqnarray}
	\op{I}-\op{W}_\mathrm{t}^\dag\op{W}_\mathrm{t}
	&=&
	\op{I}-\op{F}_N^\dag\op{W}_\mathrm{f}^\prime\op{W}_\mathrm{f}^*\op{F}_N
	\nonumber
	\\
	&=&
	\op{F}_N^\dag\op{I}\,\op{F}_N-\op{F}_N^\dag(\op{I}-\op{E}^\dagger\op{E})^*\op{F}_N
	\nonumber
	\\
	&=&
	\op{F}_N^\dag(\op{E}^\dagger\op{E})^*\op{F}_N.
	\nonumber
	\label{eq:warptimedualtail}
\end{eqnarray}
The generic $ik$-th block item of square block matrix $\op{F}_N^\dag\op{H}^\prime\op{Z}^*\op{H}^*\op{F}_N$ results
\begin{equation}\label{eq:warptimedualker}
	(\op{S}_i^*\op{V}\,\op{Q}_i^*\op{F}_N)^\dag\op{Z}^*(\op{S}_k^*\op{V}\,\op{Q}_k^*\op{F}_N)
	\nonumber
\end{equation}
where the same matrix $\op{S}_k^*\op{V}\,\op{Q}_k^*\op{F}_N$ used for aliasing in \eqref{eq:timealias} has been highlighted, meaning that $\IWT$ can be computed by using the same basis matrixes employed for aliasing compensation.

\begin{figure}[t]
	\centering%
	\includegraphics[scale=.75,trim=0 10pt 0 10pt]{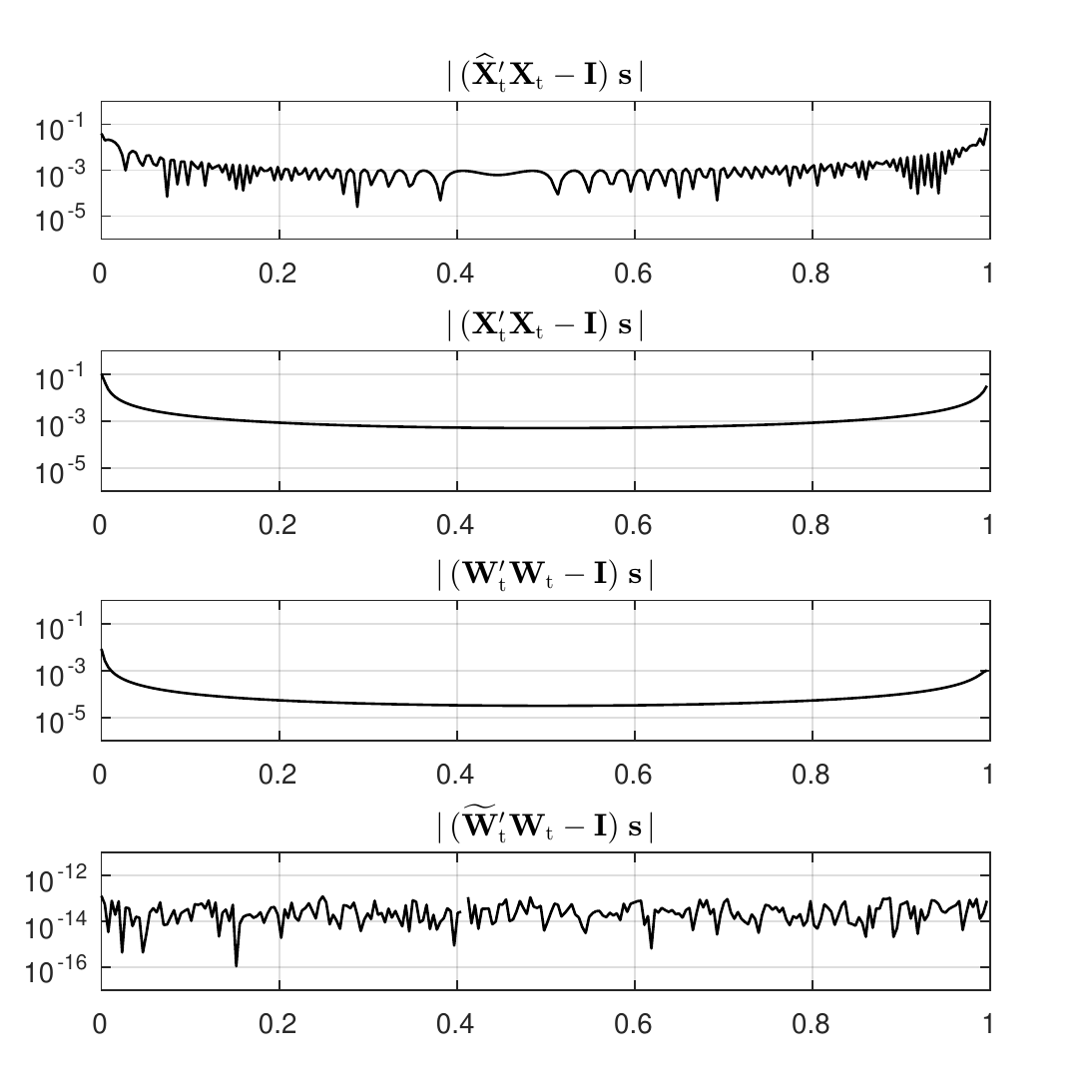}%
	\caption{%
		Reconstruction errors with respect to a random input signal corresponding to the inverse map sampled operator \eqref{eq:warptimealiasinv}, \ac{SWF} \eqref{eq:warptimealias}, \ac{SAF} \eqref{eq:warptimefromalias} and \ac{SAF} dual operator \eqref{eq:warptimedualdec}. The exponential map $w(t)=2^t-1$, $t\in[0,1]$ has been employed with $M/N\simeq{}2$.
	}%
	\label{fig:warperr}
\end{figure}

In order to exemplify the accuracy which can be achieved according to the various considered inversion strategies, reconstruction errors with respect to a random input signal have been plotted in Fig.~\ref{fig:warperr}. A complete analysis of the error behaviour should take into account parameters such as the redundancy $M/N$ (see~\cite{Caporale2009,Caporale2014}). Nevertheless, Fig.~\ref{fig:warperr} shows that $\widehat{\op{X}}^\prime_\mathrm{t}$ is about as accurate as ${\op{X}}^\prime_\mathrm{t}$, but suggests that the corresponding error matrix may not have a low rank. A thorough error analysis will be provided in Section~\ref{sec:results}.

Finally, as far as the computation is concerned, the whole \ac{FW} to \ac{TW} transition can be summarized by the following operations:
\begin{itemize}[leftmargin=14pt]
	\item replace $\op{X}_\mathrm{f}$ with $\op{X}_\mathrm{t}$
	\item replace basis vectors $\op{U}$ and $\op{V}$ with their Fourier transforms
	\item replace kernels $\op{S}$ and $\op{Z}$ with their conjugates
	\item replace modulation matrixes $\op{P}$ and $\op{Q}$ with corresponding circulant shifts.
\end{itemize}
Other issues regarding convergence and constraints on \emph{windowing/filtering} will be discussed in next subsections.

\subsection{Input and output domains indexing}
\label{subsec:freq2timedom}

The aim of this subsection is to focus on how differently the sets $\mathds{Z}_{N,L_N}$ and $\mathds{Z}_{M,L_M}$ have to be chosen in \ac{FW} and \ac{TW} with respect to the decomposition and algorithm involved in the \ac{SAF} approach. When introducing \eqref{eq:warptime} and \eqref{eq:warpfreq} and more remarkably \eqref{eq:tailsdec} and \eqref{eq:aliasdec}, constraints for invertibility and representation convergence have not been given. So, the feasibility conditions will be reviewed and updated to the generalized setting \eqref{eq:inputshift}.

A necessary condition for invertibility is expressed by
\begin{equation}\label{eq:redun}
	\frac{M}{N}>\max{}Dw.
\end{equation}
This condition can be empirically obtained by imposing that the sampling \eqref{eq:warpfreqaliaslong} of the continuous axis in \eqref{eq:warpinf} is capable of catching the band enlargement brought by warping. However, by analysing the structure of matrix $\op{W}$ (see Fig.~\ref{fig:warpalg}), it turns out that condition \eqref{eq:redun} is not sufficient, as a similar condition has to be satisfied separately for positive and negative indexes
\begin{equation}\label{eq:redunnegpos}
	\frac{M-L_M}{N-L_N}>
	\max{}Dw
	\qquad
	\frac{L_M}{L_N}>
	\max{}Dw.
\end{equation}
In previous works only the symmetrical sets $\mathds{Z}_{N,N/2}$ and $\mathds{Z}_{M,M/2}$, with $N$ and $M$ even, have been considered, hence \eqref{eq:redun} was also implying conditions \eqref{eq:redunnegpos}.

As far as \ac{FW} is concerned, it is worth highlighting how the indexing of the input domain affects the way the temporal component of the input signal are treated. From a qualitative point of view, samples corresponding to positive indexes are processed in a causal way, whereas samples corresponding to negative samples are processed in a non-causal way. So, if $L_N=0$, \ac{FW} behaves qualitatively as a causal transformation. Conversely, from a quantitative point of view, every warping map belonging to the class of piecewise smooth functions acts on positive indexes in a non-causal way because of the slow decay of $\op{W}(m,n)$ along $m$ (this is true also for smooth maps except the map $w(x)=\nicefrac{1}{\pi}\,\mathrm{atan}(\nu\mathrm{tan}(\pi{}x))$, where $\nu$ is a suitable parameter $\nu\in(0,\infty)$, being the only purely causal map). So, the non-causal behaviour is inherent with the considered warping maps and implies that the output domain has to be chosen accordingly. However, it is interesting to analyse if the proposed computational approach could be used with input and output domains as close as possible to $\mathds{Z}_{N,0}$ and $\mathds{Z}_{M,0}$.
 
As far as \ac{TW} is concerned, the input and output domains have to be chosen to guarantee that equation \eqref{eq:warptimewind} produces a real signal. The symmetrical choice $\mathds{Z}_{N,\lfloor{}N/2\rfloor}$ and $\mathds{Z}_{M,\lfloor{}M/2\rfloor}$ intuitively suits this requirements as $\op{L}_N$ and $\op{L}_M$ operate in the frequency domain, meaning that the indexes in $\mathds{Z}_{N,L_N}$ and $\mathds{Z}_{M,L_M}$ are frequencies and for each positive frequency a negative one has to be present. This and other constraints for \ac{TW} will be detailed later in this Section.

The set $\mathds{Z}_{N,\lfloor{}N/2\rfloor}$ represents a sampling over a period of the time domain for \ac{FW} and of the frequency domain for \ac{TW}. In digital signal processing a major difference holds between sampling an interval in even or odd number of points. Therefore, it is sensible to introduce the period boundaries not dependant on the particular sampling choice. To do this, the interval covered by $\mathds{Z}_{N,L_N}$ is thought at as a continuous interval of length $N$, whose left and right boundaries are
\begin{eqnarray}
	z_{N,l}
	&=&
	L_N+\nicefrac{\mathrm{mod}_2N}{2}
	\label{eq:sxbnd}
	\\
	z_{N,r}
	&=&
	N-L_N-\nicefrac{\mathrm{mod}_2N}{2}.
	\label{eq:rxbnd}
\end{eqnarray}
As instance, when $N$ is even, the $\mathds{Z}_{N,L_N}$ covers the interval $[-L_N,N-L_N)$, whereas when $N$ is odd $\mathds{Z}_{N,L_N}$ covers the interval $[-L_N-\nicefrac{1}{2},N-L_N-\nicefrac{1}{2})$. Then, the relative shift $\mu_N\in[0,1]$ representing how much the set $\mathds{Z}_{N,L_N}$ differs from the set $\mathds{Z}_{N,\lfloor{}N/2\rfloor}$ is introduced
\begin{equation}\label{eq:normshift}
	\mu_N=\frac{\max(z_{N,l},z_{N,r})-N/2}{N/2}.
\end{equation}
When $N_L$ is equal to $\lfloor{}N/2\rfloor$, $\mu_N$ is equal to $0$ for both even and odd $N$. So, $\mu_N=0$ represents the symmetric indexing case, while $\mu_N=1$ represents the causal case. We also highlight the following
\begin{eqnarray}
	\max(z_{N,l},z_{N,r})
	&=&
	(1+\mu_N)N/2
	\label{eq:rxbndnorm}
	\\
	\mathrm{min}(z_{N,l},z_{N,r})
	&=&
	(1-\mu_N)N/2.
	\label{eq:sxbndnorm}
\end{eqnarray}
So, conditions \eqref{eq:redunnegpos} can can be rewritten in the following way
\begin{equation}\label{eq:redunsxrx}
	\frac{M}{N\max{}D{w}}\frac{1+\mu_M}{1+\mu_N}>1,
	\quad
	\frac{M}{N\max{}D{w}}\frac{1-\mu_M}{1-\mu_N}>1.
\end{equation}
These conditions are necessary for both \ac{SWF} and \ac{SAF}. However, for the \ac{SAF} approach to be implementable by taking advantage of decompositions \eqref{eq:tailsdec} and \eqref{eq:aliasdec}, conditions \eqref{eq:redunsxrx} are necessary but not sufficient. To show this and obtain a sufficient condition we review and update the definition of $\op{V}$, $\op{Y}$, $\op{U}$ and $\op{S}$. Following, the approach pursued in previous works, $\op{V}$, $\op{Y}$ and $\op{U}$ are defined as sampling of continuous non-divergent functions. In more detail, the $k$-th column of matrix $\op{V}$ is obtained by sampling the polynomial $z^k$, $k\geq{}0$, defined on $[z_{N,l},z_{N,r})$, hence it must be normalized with respect to its maximum which occurs on one of the boundaries and is identified by \eqref{eq:rxbndnorm}, hence we set
\begin{equation}\label{eq:npow}
	\op{V}(k,n)=\frac{n^k}{((N/2)(1+\mu_N))^k}
	\quad
	n\in\mathds{Z}_{N,N_L}.
\end{equation}
The $i$-th row of matrix $\op{Y}$ is obtained by sampling the function $z^{-i}$, $i\geq{}0$, defined on $(-\infty,z_{M,l})\cup[z_{M,r},\infty)$, whose maximum occurs in \eqref{eq:sxbndnorm}, so it follows
\begin{equation}\label{eq:mpow}
	\op{Y}(m,i)=\frac{m^{-(i+1)}}{((M/2)(1-\mu_M))^{-(i+1)}}
	\quad
	m\notin\mathds{Z}_{M,L_M}.
\end{equation}
Finally, The $i$-th row of matrix $\op{U}$ is obtained by sampling the derivatives of the following function $\zeta$
\begin{equation}
	\zeta(z)=\pi\cot(\pi z)-\frac{1}{z}=\sum_{k\neq0}\frac{1}{z-k}
	\qquad
	|z|<1
	\label{eq:cot}
\end{equation}
where the boundaries are obtained by normalizing \eqref{eq:sxbnd}-\eqref{eq:rxbnd} and taking advantage of \eqref{eq:rxbndnorm}-\eqref{eq:sxbndnorm}. Function $\zeta$ originates from periodically summating $\op{Y}(m,0)$ according to \eqref{eq:alias}. So, we set
\begin{equation}\label{eq:mcot}
	\op{U}(m,i)=-\frac{(\mu_M-1)^{i+1}}{2^{i+1}i!}\,D^{i}\zeta(m/M)
	\quad
	m\in\mathds{Z}_{M,L_M}.
\end{equation}

Definitions \eqref{eq:npow}, \eqref{eq:mpow} and \eqref{eq:mcot} have been given in such a way that by posing $\mu_N=\mu_M=0$ one gets the definition previously given in~\cite{Caporale2009a}. Also, they generalize to the the case of odd values for $N$ or $M$.

\begin{figure}[t]
	\centering%
	\includegraphics[scale=.75,trim=0 10pt 0 10pt]{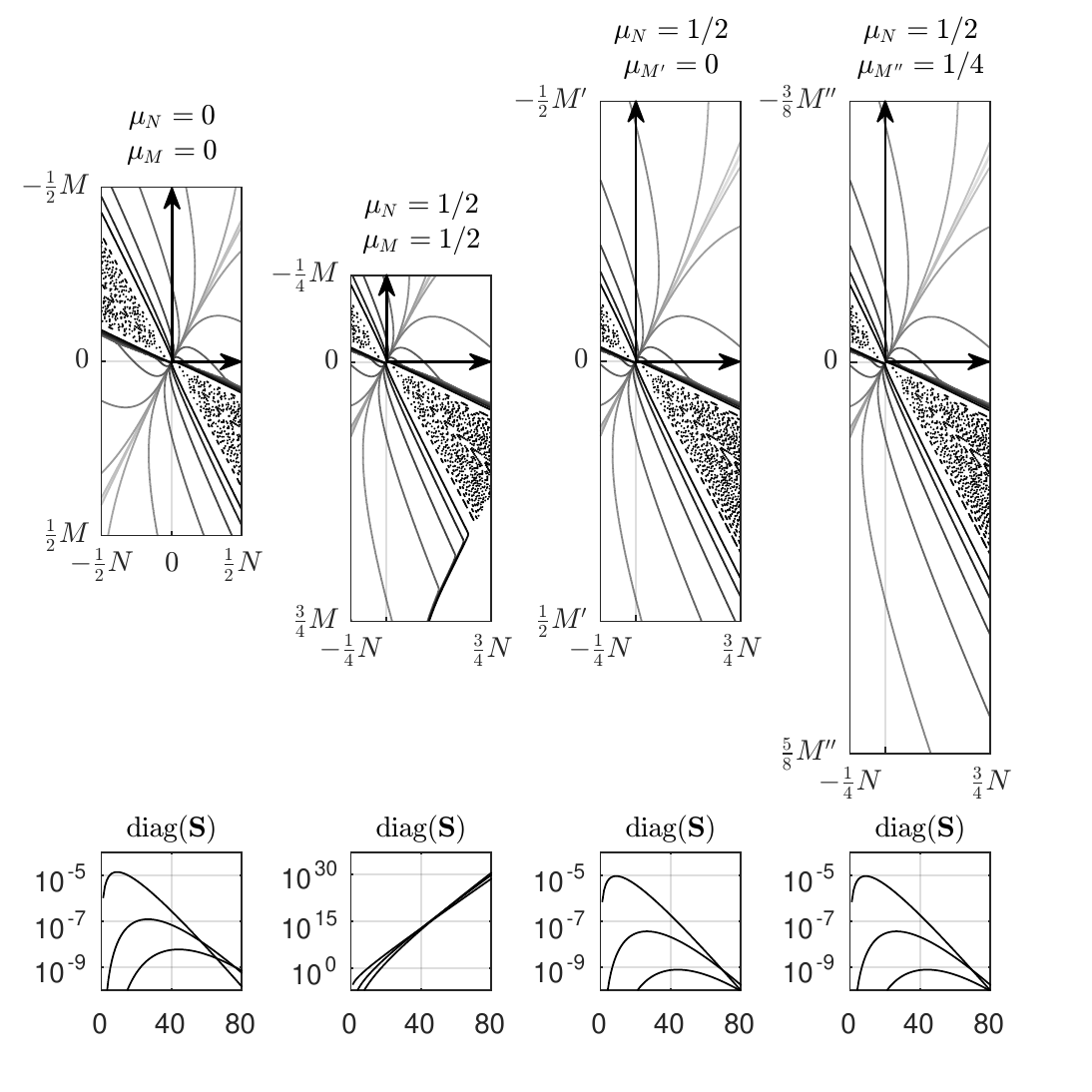}%
	\caption{%
		Input and output domain variations and impact on aliasing decomposition convergence for a \ac{FW} map having $\max{}Dw=2$. From the left to the right isolines of matrix $\op{W}_\mathrm{f}$ relative to: (i) the symmetrical case; (ii) a non-symmetrical input and output case satisfying \eqref{eq:redun} and \eqref{eq:redunnegpos} but not satisfying \eqref{eq:redundsing}, thus $\op{X}_\mathrm{f}$ can be computed correctly but $\op{W}_\mathrm{f}$ cannot be retrieved; (iii) a non-symmetrical input case having the same convergence rate as case (i) as the ratio $M'/(N(1+\mu_{N}))$ is equal to $M/N$ (represented graphically by the ratio between the arrow lengths); (iv) a non-symmetrical input and output case showing that increasing $M$ cannot be used to decrease the anti-causal samples while maintaining the same convergence rate.
	}%
	\label{fig:warpdmn}
\end{figure}

In~\cite{Caporale2009a} the kernel $\op{S}_l$ referred to a generic singularity $\xi_l$ has been also decomposed in two terms, one depending on the map only and the other one depending on the parameter setting, i.e. the relative redundancy $M/(NDw(\xi_l))$. The first term remain unchanged with respect to the generalized indexing \eqref{eq:inputshift}, whereas the second one has to be redefined in order to include the relative shifts $\mu_N$ and $\mu_M$. So, $\op{S}_l$ is given by
\begin{equation}\label{eq:kernelS}
	\op{S}_l=[\op{K}^{+}_l-\op{K}^{-}_l]\cdot{}\op{J}_l
\end{equation}
where $\cdot$ represents the element by element product, $\op{K}_l$ is a lower triangular matrix evaluated either in $\xi_l^+$ or in $\xi_l^-$ which will be detailed later
\begin{equation}\label{eq:kernelK}
	\op{K}_l(i,k)=
	\left\{%
	\begin{array}{ll}
	\alpha_{i,i-k}(\xi)	&	k\leq i
	\\
	0			&	\mathrm{otherwise}
	\\
	\end{array}
	\right.
	\quad{i},{k}=0,1,\ldots
\end{equation}
ans $\op{J}_l$ is a scaling matrix whose elements are given by
\begin{equation}\label{eq:kernelJ}
	\op{J}_l(i,k)=J(\xi_l)^{-k}(-j\pi{}M(1-\mu_M))^{k-i-1}
\end{equation}
whose decaying behavior is mainly affected by $J(\xi_l)$ 
\begin{equation}\label{eq:redunshift}
	J(\xi_l)=\frac{M}{ND{w}(\xi_l)}\frac{1-\mu_M}{1+\mu_N}.
\end{equation}
So, a necessary condition for the decompositions \eqref{eq:tailsdec} and \eqref{eq:aliasdec} to converge is $J(\xi_l)>1$, i.e.
\begin{equation}\label{eq:redundsing}
	\frac{M}{ND{w}(\xi_l)}\frac{1-\mu_M}{1+\mu_N}>1
	\qquad
	l=1,\ldots,I.
\end{equation}
When $\mu_M=\mu_N=0$, $J(\xi_l)>1$ is guaranteed by \eqref{eq:redun}. For different choices, it might be required to increase $M$. It is worth pointing out that condition \eqref{eq:redundsing} refers to the derivative in singularity points, i.e. $Dw(\xi_l), l=1,\ldots,I$ and not to $\max{}Dw$. If $\max{}Dw$ occurs in one of the singularities, then \eqref{eq:redundsing} implies both \eqref{eq:redunsxrx}.

The parameter $J(\xi_l)$, being responsible for the vanishing of $\op{S}_l$, also determines how to truncate the dimension of $\op{S}_l$ in order to obtain a numerically accurate representation. So, for $\mu_M\neq\mu_N\neq0$ a kernel $\op{S}_l$ vanishing in the same way as for $\mu_M=\mu_N=0$ can be obtained only with a larger relative redundancy $M/(NDw(\xi_l))$. As instance, with $\mu_N=\mu_M=\nicefrac{1}{2}$, the number of output samples needs to be $3$ times larger. In a real application, one can assume that $N$ and $\mu_N$ are set by specifications, while $M$ and $\mu_M$ can be adjusted in order to meet a target $J$, which implies constant $M(1-\mu_M)$. From eq. \eqref{eq:sxbndnorm} and \eqref{eq:sxbnd}-\eqref{eq:rxbnd} one gets $M(1-\mu_M)\simeq{2\mathrm{min}(L_M,M-L_M)}$. So, any choice of $M$ and $\mu_M$ leaves $\mathrm{min}(L_M,M-L_M)$ unchanged. With respect to the pure causal case $\mathds{Z}_{N,0}$ and $\mathds{Z}_{M,0}$, one would have $\mu_M\to{}0$, thus $M\to\infty$ but still $L_M>0$ and equal to the case $\mu=0$ and $M=2L_M$. These considerations are exemplified in Fig.~\ref{fig:warpdmn}, where different choices of the output domain are considered.

This short analysis clarifies the fact that for \ac{FW} there is a certain degree of freedom in the choice of the input indexing, whereas the most convenient output indexing is always the same as if the input indexing was enlarged and made symmetrical. This makes the employment of the generalized indexing quite unpractical unless computational complexity is not an issue and the focus is just on  getting a correct numerical representation of the transformation.

We now go back to \ac{TW} and the way it is conceived starting from \ac{FW}. By comparing \eqref{eq:warptime} and \eqref{eq:warpfreq}, the choice of the indexing regards the output of $\op{F}_N$, as it is fed into operator $\op{L}_M\op{W}^*\op{L}_N^\prime$, which is not shift-invariant. Hence, $\mathds{Z}_{N,L_N}$ and $\mathds{Z}_{M,L_M}$ must be chosen in such a way that if the input of $\op{L}_M\op{W}^*\op{L}_N^\prime$ is a spectrum of a real signal, then its output has to be a spectrum of a real signal too. The property is easily verified for the $\infty\times\infty$ core operator $\op{W}$. In fact, by introducing operator $\op{R}$ performing the indexing reversal
\begin{equation}\label{eq:reverse}
	\op{R}(l,k)=\delta(l+k)
	\quad
	\nonumber
\end{equation}
one has to impose that, with respect to an $\infty\times{}1$ input signal $\op{s}$ such that $\op{R}s^*=s$, the output is also invariant with respect to the application of reversing and conjugation, that is
\begin{equation}\label{eq:reversesig}
	\op{R}(\op{W}\op{s})^*=\op{W}\op{s}
	\quad
	s.t.
	\quad
	\op{R}s^*=s
	\nonumber
\end{equation}
which is verified if
\begin{equation}\label{eq:reverseinf}
	\op{W}=\op{R}\,\op{W}^*\op{R}.
	\nonumber
\end{equation}
The above condition is the same as stating that $\op{W}(m,n)=\op{W}^*(-m,-n)$, which has already pointed out. Conversely, for the truncated operator $\WT$, the condition is rewritten as
\begin{equation}\label{eq:reversewind}
	\op{L}_M^{\phantom{\prime}}\op{W}^*\op{L}_N^\prime=\op{R}_M\,\op{L}_M^{\phantom{\prime}}\op{W}\,\op{L}_N^\prime\op{R}_N
	\nonumber
\end{equation}
where $\op{R}_M$ and $\op{R}_N$ are now square matrixes of size $M$ and $N$. For this condition to be verified the following choices are required
\begin{eqnarray}
	\mathds{Z}_{N,L_N}
	&=&
	\{-(N-1)/2\ldots,(N-1)/2\}
	\quad
	\;\,N\;\mathrm{odd}
	\nonumber
	\label{eq:inputsym}
	\\
	\mathds{Z}_{M,L_M}
	&=&
	\{-(M-1)/2\ldots,(M-1)/2\}
	\quad
	M\;\mathrm{odd}.
	\nonumber
	\label{eq:outputsym}
\end{eqnarray}
Symmetry on the time domain comes trivially from the Fourier transform. Less obviously, only odd $N$ and $M$ are allowed. Critically sampled signals, i.e. $N$ even, could be still dealt with by resampling them on $N+1$ frequency points by splitting the $N/2$-th frequency coefficient over $N/2$ and $-N/2$. In a similar fashion, an even $M$ could be obtained by forcing an additional oversampling after warping.

Finally, we can summarize the results about domain constraints by the following statement. The employment of \ac{SWF} for \ac{FW} comes with the cost of employing a non strictly causal transformation. This drawback can still be overtaken in some time-frequency analysis application such as \ac{CQT}, while it represents a major limitation for the applications where \ac{FW} represents a way to compensate for physical phenomena as dispersive propagation. Conversely, the employment of \ac{SWF} for \ac{TW} comes with the minor constraints of taking only odd values for $N$ and $M$.

\section{Warping as Interpolation}
\label{sec:warp2interp}

\begin{figure}[t]
	\centering%
	\includegraphics[scale=.75,trim=0 10pt 0 10pt]{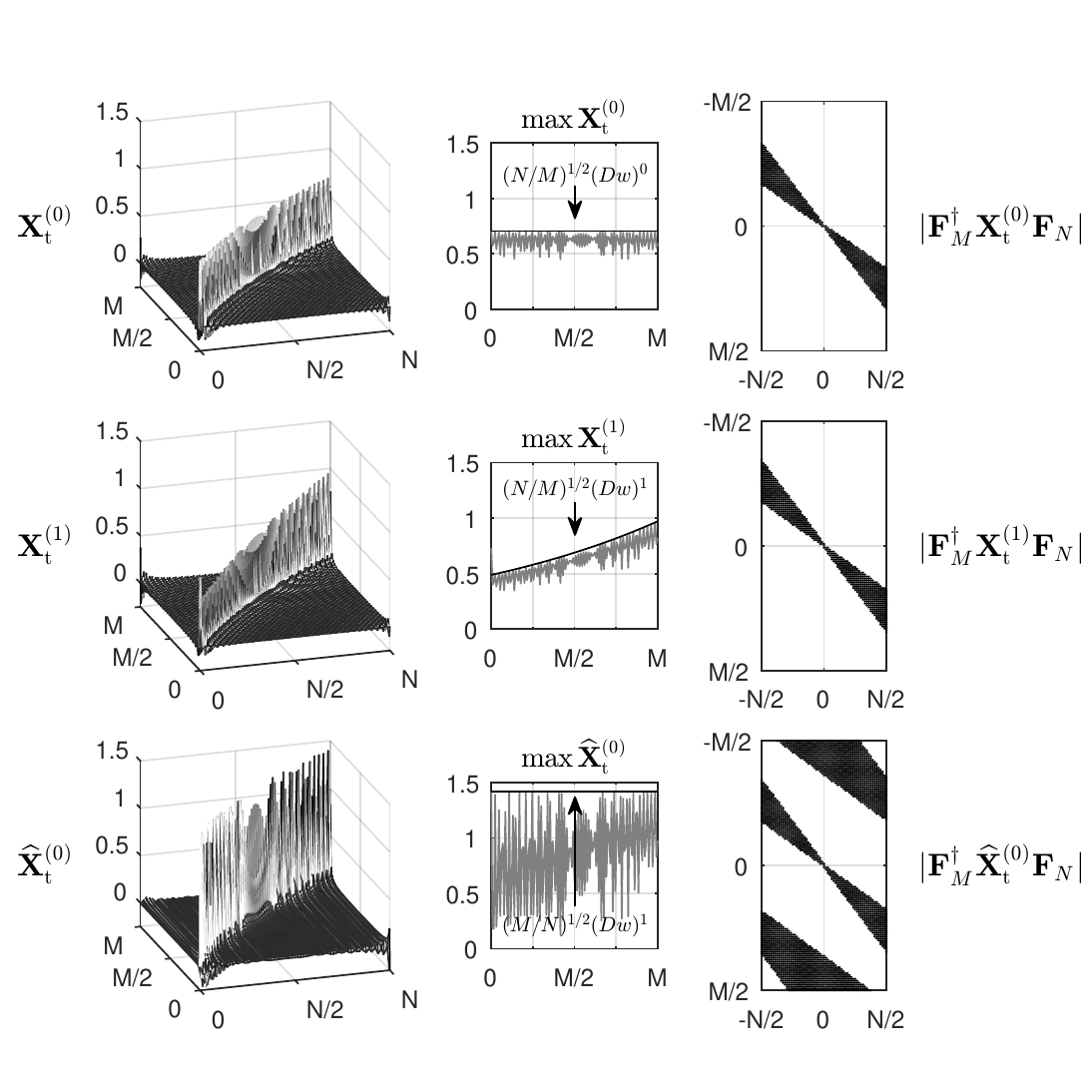}%
	\caption{%
		Time and frequency domain structure of an interpolator represented by operator $\XT^{(0)}$ and the two possible inverse operators $\XT^{(1)}$ and $\widehat{\op{X}}_\mathrm{t}^{(0)}$ obtained by taking advantage of the warping approach and of the inverse map respectively. The behavior of the operators maxima are related to the interpolation map derivatives (central column) and their sparsity patterns in the frequency domain are shown. The different sparsity pattern of $\widehat{\op{X}}_\mathrm{t}^{(0)}$ is due to aliasing along the row direction.
	}%
	\label{fig:warpsparse}
\end{figure}

As discussed in \ref{sec:intro} and \ref{subsec:contspace}, the entire design of warping operator  is driven by orthogonality, but, in certain cases, the presence of the orthogonalizing factor is not practical either for \ac{TW} and \ac{FW}, whereas the availability of the inverse operator is still necessary. The suppression of the orthogonalizing factor, i.e. considering the interpolation operator $\XT^{(0)}$, does not allow any more to employ the transpose operator for recovering the original signal and requires instead an approximation of the continuous operator $\mathfrak{W}^{(1)\dag}=\mathfrak{V}^{(0)}$, as shown in \ref{subsec:contspace}. The discretization of $\mathfrak{W}^{(1)}$ brings to $\XT^{(1)}$, whereas the discretization of $\mathfrak{V}^{(0)\dag}$ brings to $\widehat{\op{X}}_\mathrm{t}^{(0)}$. To clarify the difference between these approaches, operators $\XT^{(0)}$, $\XT^{(1)}$ and $\widehat{\op{X}}_\mathrm{t}^{(0)}$ have been depicted in Fig.~\ref{fig:warpsparse} together with their sparsity pattern in the frequency domain. In the time domain, both $\XT^{(0)}$ and $\widehat{\op{X}}_\mathrm{t}^{(0)}$, being pure interpolators, appear as a constant warped diagonal matrix. Operator $\XT^{(1)}$ differs from $\XT^{(0)}$ for an amplitude factor, hence in the frequency domain they share the same sparsity pattern. The sparsity pattern of $\widehat{\op{X}}_\mathrm{t}^{(0)}$ originates from aliasing along rows, thus limiting the reconstruction performances as it will be shown in Section~\ref{sec:results}. For this reason, we focus on extending the results obtained for $\WT^{(\nicefrac{1}{2})}$ to $\WT^{(0)}$.

With reference to \ac{SWF} for \ac{TW}, the difference between $\op{X}_\mathrm{t}^{(\nicefrac{1}{2})}$ and $\op{X}_\mathrm{t}^{(0)}$ is merely an amplitude factor. Conversely, when considering \ac{SAF}, the difference between $\op{W}_\mathrm{t}^{(\nicefrac{1}{2})}$ and $\op{W}_\mathrm{t}^{(0)}$ involves reviewing the decompositions \eqref{eq:tailsdec} and \eqref{eq:aliasdec}. From a practical point of view this can be accomplished by redefining the matrix $\op{K}_l$ \eqref{eq:kernelK} employed to express the kernel $\op{S}_l$ \eqref{eq:kernelS}.
Rather than reviewing the whole strategy for obtaining the decompositions, we just recall that the factorization model comes from the possibility of expliciting the dependency of $\op{E}_i(m,n)$ on $m$ by
\begin{equation}\label{eq:tailmod}
	\op{E}_i(m,n)=\mathrm{e}^{j2\pi{}m\xi_i}\sum_{i=0}^{\infty}{}D^i\phi_n(\xi_i^{\pm})(-j2\pi{}m)^{(i+1)}
\end{equation}
where  $D^i\phi_n(\xi_i^{\pm})$ stays for the differential value $D^i\phi_n(\xi_i^+)-D^i\phi_n(\xi_i^-)$ and $\phi_n(x)=\op{F}_w(n,x)$, that is
\begin{equation}\label{eq:func}
	\phi_n(x)=(Dw(x))^{\nicefrac{1}{2}}\,\mathrm{e}^{-j2\pi{}nw(x)}.
\end{equation}
The model \eqref{eq:tailmod} is responsible for matrix $\op{Y}$ \eqref{eq:mpow}. By expressing the derivatives $D^i\phi_n(x)$ in a symbolic way with respect to the derivative order, matrix $\op{U}$ \eqref{eq:npow} can be explicated and the value of coefficients $\alpha$ in \eqref{eq:kernelK} found. As already done in \eqref{eq:warpkernonort}, we here replace the power $\nicefrac{1}{2}$ with a generic power $b\in[0,1]$ and show that the resulting expression is compliant to the one obtained for $\nicefrac{1}{2}$. In \ref{subsec:contspace} we already pointed out that the value of $b$ being interesting from a practical point of view are $0$ and $1$. For these two values, the adopted model would not work for reasons which will explained in \ref{subsec:dernonort}. The workaround employed here consists in solving the problem symbolically also with respect to $b$. This is also the reason why we have considered the \emph{soft} orthogonalizing factor $(Dw)^b$ in \eqref{eq:warpkernonort}. Moreover, we here detail the symbolic algorithm for finding the coefficients $\alpha$. 

Before dealing with the decomposition \eqref{eq:tailmod}, we aim to show that, given the possibility of expressing $\op{W}^{(b)}$ and $\op{W}^{(1-b)}$ with a \ac{SAF} approach, then the analytical dual operators can also be defined.

\subsection{The generalized dual operator}
\label{subsec:inversenonort}

As a starting point, the basic operator $\op{W}$ in \eqref{eq:warpinf} is redefined by changing the exponent of the orthogonalization factor from $\nicefrac{1}{2}$ to a generic exponent $b\in[0,1]$
\begin{equation}\label{eq:warpinfnonort}
	\op{W}^{(b)}(m,n)=\int_0^1(Dw(x))^b\,e^{j2\pi(mf-nw(x))}dx.
\end{equation}
All related operators such as $\op{A}$ and $\op{E}$ will be referred to in the same way. As shown in \eqref{eq:warpnonortid}, inversion can be still obtained by applying operator $\op{W}_{\bar{b}}$ where $\bar{b}=1-b$
\begin{equation}\label{eq:warpinfnonortid}
	\op{W}^{(\bar{b})\dag}\op{W}^{(b)}=\op{I}.
	\nonumber
\end{equation}
By restricting the input by $\op{L}_N^{\prime}$ and employing the splitting between $\WF$ and $\op{E}$ as done in \eqref{eq:tails} we get
\begin{equation}\label{eq:warpfreqnonortid}
	(\WF^{(\bar{b})}+\op{E}^{(\bar{b})})^\dag(\WF^{(b)}+\op{E}^{(b)})=\WF^{(\bar{b})\dag}\WF+\op{E}^{(\bar{b})\dag}\op{E}^{(b)}=\op{I}.
	\nonumber
\end{equation}
which can be used as in \eqref{eq:warpfreqdual}. It is worth noting that
\begin{eqnarray}
	\WF^{(\bar{b})\dag}\WF^{(b)}
	&\neq&
	\WF^\dag\WF
	\nonumber
	\label{eq:warpwindnoneq}
	\\
	\op{E}^{(\bar{b})\dag}\op{E}^{(b)}
	&\neq&
	\op{E}^\dag\op{E}
	\nonumber
	\label{eq:warptailnoneq}
\end{eqnarray}
because truncations are operated separately, nevertheless $\WF^{(\bar{b})\dag}\WF^{(b)}$ is still an approximation of the identity as accurate as $\WF^\dag\WF$, hence its inverse can still be calculated by means of the Neumann series
\begin{equation}\label{eq:eq:warpfreqnonortdual}
	\IWF^{(\bar{b})}=\WF^{(b)}\left(\op{I}+\sum_{k=1}^{\infty}(\op{E}^{(\bar{b})\dag}\op{E}^{(b)})^k\right).
	\nonumber
\end{equation}
Most importantly, in \ref{subsec:dernonort} and \ref{subsec:algnonort} it will be shown that $\op{E}^{(b)}$ and $\op{E}^{(\bar{b})}$ share the same decomposition as $\op{E}^{(b)}$ reported in \eqref{eq:tailsdec} but with different kernels. As a consequence, equation \eqref{eq:warpfreqdualdec} can be rewritten by just suitably replacing $\op{H}$ with either $\op{H}^{(b)}$ or $\op{H}^{(\bar{b})}$ and kernel $\op{Z}$ with $\op{Z}^{(\bar{b},b)}$, which is properly recomputed by using the pair $\op{S}_i^{(\bar{b})}$ $\op{S}_k^{({b})}$:
\begin{equation}\label{eq:eq:warpfreqnonortdec}
	\IWF^{(\bar{b})}=\WF^{(b)}(\op{I}+\op{H}^{(\bar{b})\dag}\op{Z}^{(\bar{b},b)}\op{H}^{(b)}).
	\nonumber
\end{equation}
Finally, the extension to \ac{TW} is simply obtained by applying the same procedure as in \ref{subsec:freq2timeinv}, i.e.
\begin{equation}\label{eq:eq:warptimenonortdec}
	\IWT^{(\bar{b})}=\WT^{(b)}(\op{I}+\op{F}_N^\dag\op{H}^{(\bar{b})\prime}\op{Z}^{(\bar{b},b)*}\op{H}^{(b)*}\op{F}_N).
\end{equation}
The computation of $\op{Z}^{(\bar{b},b)}$ can be easily inferred by~\cite{Caporale2014}.

\subsection{Polynomials identification}
\label{subsec:dernonort}

The problem of expressing the derivative of the composition of two functions, being here the exponential function and the warping function with some additional coefficients and factors as represented in \eqref{eq:tailmod} and \eqref{eq:func}, has a notable solution known as Fa\`{a} di Bruno's formula, which also has a simpler form in case the first function is the exponential. Nevertheless, to the best of our knowledge the decomposition proposed here is remarkably different since the inherent recursive structure of the exponential function derivatives is exploited. Moreover, the proposed approach is more effective from the algorithmic point of view with respect to our purposes. In fact, we specifically aim to express the derivative of \eqref{eq:func} as a power series of $-j2\pi{}nDw(x)$, which allows for the decomposition \eqref{eq:tailsdec} and for an effective characterization of convergence~\cite{Caporale2014}. We also point out that this symbolic computation cannot be performed by any of the symbolic softwares available.

Given the following generalization of functions \eqref{eq:func}
\begin{equation}\label{eq:funcnonort}
	\phi^{(b)}_a=\mathrm{e}^{aw}\,(Dw)^b
\end{equation}
where the derivative is intended with respect to the implied variable $x$ and $a=-j2\pi{}n$, the $k$-th derivative can be expressed as follows
\begin{equation}\label{eq:derfuncnonort}
	D^k\phi_{a}^{(b)}=\mathrm{e}^{aw}\sum_{l=0}^{k}\alpha_{k,k-l}^{(b)}(aDw)^l
\end{equation}
where the coefficients $\alpha_{k,l}^{(b)}$ can be iteratively obtained by the following iterative relationships
\begin{eqnarray}
	\alpha_{k,0}^{(b)}\!
	&=&\!
	(Dw)^b
	\label{eq:derfuncinit}
	\\
	\alpha_{k+1,k+1}^{(b)}\!
	&=&\!
	D\alpha_{k,k}^{(b)}
	\label{eq:derfunciter}
	\\
    \alpha_{k+1,l+1}^{(b)}\!-\!\alpha_{k,l+1}^{(b)}\!
    &=&\!
    D\alpha_{k,l}^{(b)}\!+\!(k-l)\alpha_{k,l}^{(b)}D^2w(Dw)^{-1}
	\label{eq:derfuncdiff}
\end{eqnarray}
with $l=0,\ldots\,k-1$. \appref{The demonstration is given in Appendix \ref{app:derfuncdec}.} Equation \eqref{eq:derfunciter} is used to generate the coefficient referred to $(aDw)^0$, equation \eqref{eq:derfuncdiff} is used to extrapolate the behaviour of the coefficients $\alpha_{k,l}^{(b)}$ with respect to $k$ and equation \eqref{eq:derfuncinit} serves as initial condition. The relationship \eqref{eq:derfuncdiff} also shows that the finite difference $\alpha_{k+1,l+1}^{(b)}-\alpha_{k,l+1}^{(b)}$ is affine with respect to $(k-l)$, then it can be inferred that $\alpha_{k,l}^{(b)}$ must have a polynomial expression with respect to $k$. \appref{This implication is properly proved in Appendix~\ref{app:derfunciter}.}

With reference to the case $b=0$, we highlight the following point. If $\phi_a^{(0)}$ is considered rather than the \emph{symbolic} $\phi_a^{(b)}$
\begin{eqnarray}
	D^1\phi_a^{(0)}
	&=&
	\mathrm{e}^{aw}\left[\,aDw\,\right]
	\nonumber
	\label{eq:derfuncb01}
	\\
	D^2\phi_a^{(0)}
	&=&
	\mathrm{e}^{aw}\left[\,aD^2w+(aDw)^2\,\right]
	\nonumber
	\label{eq:derfuncb02}
\end{eqnarray}
we notice that the term in $(aDw)^0$ is not present in neither $D^1\phi_a^{(0)}$ or $D^2\phi_a^{(0)}$ and it can be inferred that is not present in any derivative. The term in $(aDw)^1$ is not present but can be forced by replacing $aD^2w$ with $D^2w(Dw)^{-1}(aDw)$. As a consequence, equation \eqref{eq:derfunciter} would not hold and the main equation for generating the new functions of the decomposition would be obtained by evaluating equation \eqref{eq:derfuncdiff} for $k=l+1$
\begin{equation}\label{eq:derfuncb0iter}
	\alpha_{k+1,k}^{(0)}
	=
	D\alpha_{k,k-1}^{(0)}+\alpha_{k,k-1}^{(0)}D^2w(Dw)^{-1}
	\nonumber
\end{equation}
with $\alpha^{(0)}_{1,0}=1$ and $\alpha^{(0)}_{2,1}=D^2w(Dw)^{-1}$. So, in principle, the algorithm could be redesigned to work for $b=0$. Conversely, with the symbolic exponent $b$, the first derivative is 
\begin{eqnarray}
	D^1\phi^{(b)}_a
	&=&
	\mathrm{e}^{aw}\,D(Dw)^b+\phi^{(b)}_a(aDw)
	\nonumber
	\\
	&=&
	\mathrm{e}^{aw}\left[\,b(Dw)^{b-1}D^2w+(Dw)^b(aDw)\right]
	\label{eq:derfunc1}
\end{eqnarray}
hence, when $b=0$, $\alpha_{1,1}^{(0)}$ is present but multiplied to a null coefficient. For the second derivative, new functions are generated by deriving $\alpha_{1,1}^{(b)}=D(Dw)^b$ while the derivation of $(Dw)^b(aDw)$ is partly explained by $D(Dw)^b$ itself:
\begin{multline}
	D^2\phi^{(b)}_a=\mathrm{e}^{aw}\Big[D^2(Dw)^b+\big[D(Dw)^b+
	\\
	(Dw)^b(Dw)^{-1}D^2w\big](aDw)\Big]+D\phi^{(b)}_a(aDw)
	\label{eq:derfunc2}
\end{multline}
Clearly, expanding the recursive term $D\phi^{(b)}_a(aDw)$ would get the manual computation infeasible after some other iterations. It is also important to highlight the effectiveness of the representation trick $(Dw)^{-1}D^2w(aDw)$ which has been employed to adhere to the power series structure. For instance, in this case, it allows to couple the term $(Dw)^b(Dw)^{-1}D^2w$ to $D(Dw)^b$. We also point out that having the coefficient $b=0$ in $D^1\phi^{(0)}_a$ does not mean that the term does not contribute the the solution. To show this concept, as an exercise, we find coefficients $\alpha_{k,1}^{(b)}$, hence we solve with respect to the factor $(Dw)^{b-1}D^2w$. By looking at $\phi^{(b)}_a$ \eqref{eq:funcnonort}, $D\phi^{(b)}_a$ \eqref{eq:derfunc1} and $D^2\phi^{(b)}_a$ \eqref{eq:derfunc2}, $(Dw)^{b-1}D^2w$ is associated to coefficients $0$, $b$ and $(2b+1)$ respectively. By inferring that for $D^k\phi^{(b)}_a$ the term $(Dw)^{b-1}D^2w$ is multiplied to $(aDw)^k$ times a $2$-nd degree polynomial, the polynomia coefficients can be calculated according to its samplings for $k=0,1,2$, thus getting $\nicefrac{1}{2}k^2+(b-\nicefrac{1}{2})k$.

In the next subsection we will prove that this methodology can be employed to obtain all coefficients $\alpha$ with respect to the general exponent $b$. Since the method would not make sense for $b=0$ and  also for $b=1$ (although it has been shown for $b=0$ only), those cases can be thought at as valid for $b\to{}0$ and $b\to{}1$.

\subsection{Polynomials computation}
\label{subsec:algnonort}

The theory for solving the decomposition \eqref{eq:derfuncnonort} has been set in~\cite{Caporale2009a}\appref{, but no proper proofs have been provided about its theoretical correctness. Moreover, a practical algorithm has not been described}. Here we provide an effective computational approach\appref{ supported by mathematical demonstration which will be detailed in Appendix \ref{app:derfuncdec}}. The dependency of coefficients $\alpha^{(b)}$ on the inherent variable $x$ and the order of derivation can be decoupled as follows
\begin{equation}\label{eq:derfuncdec}
	\alpha_{k,l}^{(b)}(x)=\sum_{n=1}^{|\Omega_l|}\beta_{l,n}^{(b)}(x)\gamma_{l,n}^{(b)}(k)
\end{equation}
where $|\Omega_l|$ is the cardinality of the set $\Omega_l=\{p_{l,1},p_{l,2},\ldots\}$ whose items are all the possible sequences satisfying
\begin{equation}\label{eq:derfuncpow}
	\sum_{m=1}^{k+1}p_{k,n,m}m=k
	\qquad
	p_{k,n,m}\left\{
	\begin{array}{cc}
	\in\mathds{Z}_- & m=1\\
	\in\mathds{Z}_+ & m>1\\
	\end{array}\right.
\end{equation}
with $\mathds{Z}_-$ and $\mathds{Z}_+$ including $0$. Functions $\beta_{l,n}$ are given by
\begin{equation}\label{eq:derfuncprod}
	\beta^{(b)}_{l,n}=(Dw)^{b}\prod_{m=1}^{l+1}(D^mw)^{p_{l,n,m}}.
\end{equation}
and  $\gamma_{l,n}(k)$ in $k$ are polynomials
\begin{equation}\label{eq:derfuncpoly}
	\gamma^{(b)}_{l,n}(k)=\sum_{m=0}^{2l}c^{(b)}_{l,n,m}k^m.
\end{equation}

\ifthenelse{\boolean{\addnum}}{
To clarify how \eqref{eq:derfuncdec}-\eqref{eq:derfuncpoly} follow from equation \eqref{eq:derfuncnonort},  by taking advantage of equations \eqref{eq:derfuncinit}-\eqref{eq:derfuncdiff}, we review the result provided at the end of \ref{subsec:dernonort} by which we found $\alpha_{k,1}^{(b)}=(Dw)^{b-1}D^2w\,(\nicefrac{1}{2}k^2+(b-\nicefrac{1}{2})k).$ As initial condition, we assume from \eqref{eq:derfuncinit} that $\beta^{(b)}_{0,1}=(Dw)^b$, $p_{0,1,1}=0$ and $\gamma^{(b)}_{0,1}=1$. From \eqref{eq:derfunciter} we get
\begin{equation}\label{eq:derfuncdec11}
	\alpha_{1,1}^{(b)}=D\alpha_{0,0}^{(b)}=D(Dw)^b=b(Dw)^{b-1}D^2w
	\nonumber
\end{equation}
hence $\beta^{(b)}_{1,1}=(Dw)^{b-1}D^2w$, $p_{1,1}=[-1,1]$ and $\gamma^{(b)}_{1,1}$ has to be determined. By employing \eqref{eq:derfuncdiff} with $l=0$ we get
\begin{eqnarray}
	\Delta_k\alpha_{k+1,1}^{(b)}
	&=&
	D\alpha_{k,0}^{(b)}+k\alpha_{k,0}^{(b)}D^2w(Dw)^{-1}
	\nonumber
	\\
	&=&
	(b+k)(Dw)^{b-1}D^2w
	\nonumber
	\label{eq:derfuncdecdiff1}
\end{eqnarray}
where $\Delta_k$ represents the finite difference operator applied to $\alpha^{(b)}_{k+1,l+1}$ with respect to $k$. The equation above states that $\Delta_k\alpha_{k+1,1}^{(b)}$ can be split in two factors being a function of $x$ and a polynomial in $k$. The polynomial $b+k$ is simply described by the coefficients $[b\quad{}1]$, while the inverse operator of the finite difference operator is
\begin{equation}\label{eq:finitesum}
	\Gamma=
	\begin{bmatrix}
	0 & \phantom{-}0 \\
	1 & -\nicefrac{1}{2}\\
	0 & \phantom{-}\nicefrac{1}{2}
	\end{bmatrix}
	\nonumber
\end{equation}
so by $\Gamma\,[b\quad{}1]^T$ we get $c^{(b)}_{1,1}=[0\quad{}b-\nicefrac{1}{2}\quad{}\nicefrac{1}{2}]$. The forward iterations require a more complex index manipulation, since, as instance
\begin{equation}\label{eq:derfuncdec22}
	\alpha_{2,2}^{(b)}=D\alpha_{1,1}^{(b)}=(b-1)(Dw)^{b-2}(D^2w)^2+(Dw)^{b-1}D^3w
	\nonumber
\end{equation}
involves two terms rather than one and only one of them is also obtained as $\alpha_{2,2}^{(b)}\,(Dw)^{-1}D^2w$.
}{}

Given the definitions \eqref{eq:derfuncdec}-\eqref{eq:derfuncpoly}, the algorithm can be split in two main parts being (i) calculation of functions $\beta_{l,n}^{(b)}$ and (ii) calculation of polynomials $\gamma_{l,n}^{(b)}$. It is worth pointing out that for representing $\beta^{(b)}_{l,n}$ and $\gamma^{(b)}_{l,n}$ we just need to manipulate the sequences $p_{l,n,m}$ and $c^{(b)}_{l,n,m}$ respectively, thus making the dependency on both $x$ and $k$ symbolic.

The functions $\beta^{(b)}_{l,n}$ could be found numerically by constrained optimization on property \eqref{eq:derfuncpow}, however an iterative computation is more practical for coupling $\beta^{(b)}_{l}$ and $\gamma^{(b)}_{l}$ correctly. First we define the following:
\begin{equation}\label{eq:gener}
	m\in\Lambda_{l,n}
	\quad
	\Leftrightarrow
	\quad
	p_{l,n,m}-b\delta_{m-1}.
\end{equation}
The sets $\Lambda_{l,n}$ will be referred to as \emph{generators}, as they identify the factors in functions $(D^mw)^{p_{l,n,m}}$ which are capable of generating an item by derivation. We also define
\begin{equation}\label{eq:powupdate}
	g_{l,n,q,m}=p_{l,n,m}-\delta_{m-\Lambda_{l,n}(q)}+\delta_{m-\Lambda_{l,n}(q)-1}	
\end{equation}
with $q\in\Lambda_{l,n}$, the following index expansion
\begin{equation}\label{eq:idxexp}
	q=\Phi_l(i),\,n=\Psi_l(i)
	\quad
	s.t.
	\quad
	i=q+\sum_{m=1}^{n-1}|\Lambda_{l,m}|
\end{equation}
and the set of cardinality $|\tilde{\Omega}_{l+1,n}|=\sum_{i=1}^{|\Omega_{l,i}|}|\Lambda_{l,n}|$ of $\tilde{p}_{l+1,n}$ sequences having length $l+2$
\begin{equation}\label{eq:powupdateidx}
	\tilde{p}_{l+1,n,m}=g_{l,\Psi_l(n),\Phi_l(n),m}
\end{equation}
which differs from the set of $p_{l+1,n,m}$ because it can feature repeated sequences. Hence, we introduce
\begin{equation}\label{eq:powupdateidxrep}
	\tilde{p}_{l+1,n,m}
	=
	p_{l+1,\Xi_l(n),m^{\phantom{-1}}}
	\qquad
	n=1,\ldots\,|\tilde{\Omega}_{l+1}|
\end{equation}
where $\Xi_l$ is non-injective so its inverse can be defined in multiple ways without affecting the result. Finally we introduce the following partitioning identifying the elements of $\Xi_l$ having the same value
\begin{equation}\label{eq:idxpart}
	i\in\Upsilon_{l,n}
	\,
	\Leftrightarrow
	\,
	\Xi_{l}(i)=n
	\qquad
	n=1,\ldots,|{\Omega}_{l+1}|
\end{equation}
such that
\begin{equation}\label{eq:pownext}
	\tilde{p}_{l+1,\Upsilon_{l,n}(q),m}=p_{l+1,n,m}
	\qquad
	q=1,\ldots,|\Upsilon_{l,n}|.
\end{equation}

By assuming $|\Omega_0|=1$, $\beta^{(b)}_{0,1}=(Dw)^b$, given the set of functions $\beta^{(b)}_{l}$, the set of functions $\beta^{(b)}_{l+1}$ can be obtained by the following procedure:
\begin{itemize}[leftmargin=14pt]
	\item find the \emph{generators} $\Lambda_{l,n}$
	\item set a $(l+2)\times|\tilde{\Omega}_{l+1,n}|$ matrix and fill it with $\tilde{p}_{l+1,n,m}$ 
	\item find repetition in $\tilde{p}_{l+1,n,m}$, thus obtain $\Xi_l(n)$ and ${p}_{l+1,n,m}$
\end{itemize}
\ifthenelse{\boolean{\addnum}}{%
As instance, $p_{0}=[0]$ with $\Lambda_{0,1}=\{1\}$, $p_{1}=[-1\quad{}1]^T$ with $\Lambda_{1,1}=\{1\quad{}2\}$, then
\begin{equation}
	p_{2}=
	\begin{bmatrix}
	p_{1,1} & p_{1,1}\\
	0 & 0
	\end{bmatrix}
	+
	\begin{bmatrix}
	-1 & \phantom{-}0\phantom{-}\\
	\phantom{-}1 &           -1\phantom{-}\\
	\phantom{-}0 & \phantom{-}1\phantom{-}
	\end{bmatrix}
	=
	\begin{bmatrix}
	-2 &           -1\phantom{-}\\
	\phantom{-}2 & \phantom{-}0\phantom{-}\\
	\phantom{-}0 & \phantom{-}1\phantom{-}
	\end{bmatrix}
	\nonumber
	\end{equation}
	with $\Lambda_{2,1}=\{1\quad{}2\}$ and $\Lambda_{1,2}=\{1\quad{}3\}$, then
	\begin{equation}
	\tilde{p}_{3}=
	\begin{bmatrix}
	p_{2,1} & p_{2,1} & p_{2,2} & p_{2,2} \\
	0 & 0 & 0 & 0\\
	\end{bmatrix}
	+
	\begin{bmatrix}
	-1 & \phantom{-}0 &           -1 & \phantom{-}0\phantom{-}\\
	\phantom{-}1 &           -1 & \phantom{-}1 & \phantom{-}0\phantom{-}\\
	\phantom{-}0 & \phantom{-}1 & \phantom{-}0 &           -1\phantom{-}\\
	\phantom{-}0 & \phantom{-}0 & \phantom{-}0 & \phantom{-}1\phantom{-}
	\end{bmatrix}
	\nonumber
\end{equation}
resulting in 
\begin{equation}
	p_{3}=
	\begin{bmatrix}
	-3 &           -2 &            -1\phantom{-}\\
	\phantom{-}3 & \phantom{-}1 &  \phantom{-}0\phantom{-}\\
	\phantom{-}0 & \phantom{-}1 &  \phantom{-}0\phantom{-}\\
	\phantom{-}0 & \phantom{-}0 &  \phantom{-}1\phantom{-}
	\end{bmatrix}
	\nonumber
\end{equation}
with $\Xi_2=[1\quad{}2\quad{}2\quad{}3]$. \appref{An accurate proof of this procedure supported by theoretical considerations on the uniqueness of the representation \eqref{eq:derfuncpow} is given in Appendix \ref{subapp:derfuncdecfunc}.}
}{}

As far as $\gamma^{(b)}_{l,n}$ is concerned, the computation is more complicated as it involves the usage of the finite difference equation \eqref{eq:derfuncdiff}. We first set
\begin{equation}\label{eq:coefupdate}
	r_{l,n,q}(k)=p_{l,n,\Lambda_{l,n}(q)}+(b+k-l)\delta_{\Lambda_{l,n}(q)-1}
	\nonumber
\end{equation}
being a polynomial in $k$ of degree $1$, whose coefficient will be referred to as $u_{l,n,q}$. The following product between polynomials
\begin{equation}\label{eq:coefupdateconv}
	\tilde{\gamma}^{(b)}_{l+1,n}(k)=\gamma^{(b)}_{l,\Psi_l(n)}(k)\,r_{l,\Psi_l(n),\Phi_l(n)}(k)
\end{equation}
will be represented by coefficients $\tilde{u}_{l+1,n}$. The following sum of polynomials
\begin{equation}\label{eq:coefupdateconvdiff}
	\Delta_k\gamma^{(b)}_{l+1,n}(k)
	=
	\sum_{q=1}^{|{\Upsilon}_{l,n}|}\tilde{\gamma}^{(b)}_{l+1,\Upsilon_{l,n}(q)}(k)
\end{equation}
\nonnum{where $\Delta_k$ represents the finite difference operator applied to $\alpha^{(b)}_{k+1,l+1}$ with respect to $k$, }
will be represented by coefficients $\tilde{c}_{l+1,n}$. Finally, assuming $\gamma_{0,1}^{(b)}=1$, given the set of polynomials $\gamma_{l}^{(b)}$, the set of polynomials $\gamma_{l+1}^{(b)}$ can be computed as follows:
\begin{itemize}[leftmargin=14pt]
	\item compute $r_{l,n,q}$ as a two element vector of coefficients $u_{l,n,q}$
	\item form a $\tilde{\Omega}_{l+1}\times{}2$ matrix having $u_{l,\Psi_l(n),\Phi_l(n),m}$ as rows
	\item form the $\tilde{\Omega}_{l+1}\times{}(2l+1)$  matrix having $c_{l,\zeta_l(n)}$ as rows
	\item perform the convolution between $u_{l,\Psi_l(n),\Phi_l(n)}$ and $c_{l,\Psi_l(n)}$
	thus obtaining the $\tilde{\Omega}_{l+1}\times{}(2l+2)$ having $\tilde{u}_{l+1,n}$ as rows
	\item identify the partitioning $\Upsilon_{l,n}$ and sum the rows belonging to the same partition, thus obtaining $\tilde{c}_{l+1,n}$
	\item apply the discrete summation operator $\Gamma$ (which can be obtained as a Pascal matrix whose rows are scaled by Bernoulli numbers).
\end{itemize}
\ifthenelse{\boolean{\addnum}}{%
As instance, $u_{0,1,1}=[b\quad{}1]$, steps $2-5$ can be skipped for $l=1$, hence $c_{1,1}=[0\quad{}b-\nicefrac{1}{2}\quad{}\nicefrac{1}{2}]$. Going forward
\begin{equation}
	u_{1,\Psi_1(n),\Phi_1(n)}
	=
	\begin{bmatrix}
	b-2 && 1 \\
	1 && 0
	\end{bmatrix}
	\quad
	c_{1,\Psi_1(n)}
	=
	\begin{bmatrix}
	c_{1,1} \\
	c_{1,1}
	\end{bmatrix}
	\nonumber
\end{equation}
the convolution gives
\begin{equation}
	\tilde{u}_{2,n}
	=
	\begin{bmatrix}
	0 && b^2 -\nicefrac{5}{2}b + 1 && \nicefrac{3}{2}b-\nicefrac{3}{2} && \nicefrac{1}{2}\\
	0 && b-\nicefrac{1}{2} && \nicefrac{1}{2} && 0
	\end{bmatrix}
	\nonumber
\end{equation}
while step $5$ can be skipped because $\Xi_1$ does not have any repeated elements, hence $\tilde{c}_{2,n}=\tilde{u}_{2,n}$ and finally operator $\Gamma$ is applied thus obtaining $c_{2,n}$
\begin{equation}
	c_{2,n}
	=
	\begin{bmatrix}
	0 & -\nicefrac{b^2}{2} + \nicefrac{3}{2}b - \nicefrac{3}{4} & \nicefrac{b^2}{2} - 2b + \nicefrac{11}{8} & \nicefrac{1}{2}b - \nicefrac{3}{4} & \nicefrac{1}{8} \\
	0 & -\nicefrac{1}{2}b+\nicefrac{1}{3} & \nicefrac{1}{2}b-\nicefrac{1}{2} & \nicefrac{1}{2} & 0
	\end{bmatrix}
	\nonumber
\end{equation}
In the above example, the index $n$ refers to the row while the column index is implied. \appref{The proof supporting this procedure is given in \ref{subapp:derfuncdecpoly}.}
}{}

\ifthenelse{\boolean{\addtab}}{
Functions $\beta^{(0)}_{l,n}$ and $\beta^{(1)}_{l,n}$ and the corresponding polynomials $\gamma^{(0)}_{l,n}$ and $\gamma^{(1)}_{l,n}$ for some values of $l$ and $n$ are given in Tables \ref{tab:betagammab0} and \ref{tab:betagammab1} respectively.

\begin{table}
	\centering
	\caption{%
		Expressions of $\beta^{(b)}_{l,n}$ and $\gamma^{(b)}_{l,n}$ for $l=0,\ldots\,3$ and $b=0$.
	}%
	\begin{tabular}{p{.2cm}p{2cm}l}
		\toprule
		$l,n$ & $\beta^{(0)}_{l,n}$	& $\gamma^{(0)}_{l,n}(k)$\\
		\midrule
		$0,1$ & $1$					& $1$\\
		\midrule
		$1,1$ & $(Dw)^{-1}D^2w$		& $\frac{1}{2}(k^2-k)$\\
		\midrule
		$2,1$ & $(Dw)^{-2}(D^2w)^2$	& $\frac{1}{8}(k^4-6k^3+11k^2-6k)$\\
		\midrule
		$2,2$ & $(Dw)^{-1}D^3w$ 	& $\frac{1}{6}(k^3-3k^2+2k)$\\
		\midrule
		$3,1$ & $(Dw)^{-3}(D^2w)^3$	& $\frac{1}{48}(k^6\!-\!15k^5\!+\!85k^4\!-\!225k^3\!+\!274k^2\!-\!120k)$\\
		\midrule
		$3,2$ & $(Dw)^{-2}D^2wD^3w$	& $\frac{1}{12}(k^5-10k^4+35k^3-50k^2+24k)$\\
		\midrule
		$3,3$ & $(Dw)^{-1}D^4w$		& $\frac{1}{24}(k^4-6k^3+11k^2-6k)$\\
		\bottomrule
	\end{tabular}
	\label{tab:betagammab0}
\end{table}
}{}

\ifthenelse{\boolean{\addpar}}{
\subsection{Further considerations on the non orthogonal case}
\label{subsec:polynonort}

We now want to focus on an aspect of the proposed approach. As highlighted in \ref{subsec:dernonort}, if $b=0$ or $b=1$ the decomposition \eqref{eq:derfuncnonort} cannot be built. For this reason, the procedure shown in \ref{subsec:algnonort} has to be intended with $b\to{}0$ and $b\to{}1$. One intuitive reason relies on the fact the decomposition accounts on negative powers of $Dw$ (see equation \eqref{eq:derfuncprod}), whereas negative powers do not raise for $b=0$ or $b=1$. Hence, imposing that the negative powers are associated with zero coefficients could be an alternative way to find the polynomials $\gamma^{(b)}_{l,n}$.

Given the following manipulation of $D^k\phi_a^{(b)}$
\begin{multline}
	D^k\phi_a^{(b)}=
	\mathrm{e}^{aw}
	\sum_{l=0}^k\sum_{n=1}^{|\Omega_{l}|}\gamma^{(b)}_{l,n}(k)a^{k-l}
	\\
	(Dw)^{k+b-l+p_{l,n,1}}\prod_{m=2}^{l+1}(D^mw)^{p_{l,n,m}}
	\nonumber
	\label{eq:derfunczeros}
\end{multline}
when $b=0$ or $b=1$, negative powers of $Dw$ must be nullified by the polynomials $\gamma^{(b)}_{l,n}(k)$. Then we have the following condition
\begin{equation}\label{eq:derfunczeroposb0}
	k<l-p_{l,n,1}-b,k\in\mathds{Z}
	\quad
	\Rightarrow
	\quad
	\gamma_{l,n}(k)=0
	\nonumber
\end{equation}
so, for $b=0$, $\gamma^{(0)}_{l,n}(k)$ has $l-p_{l,n,1}$ zeros placed on $0,\ldots,l-p_{l,n,1}-1$. Moreover, from the proposed algorithm it is easily inferred that the order of polynomial $\gamma^{(0)}_{l,n}(k)$ is equal to $l-p_{l,n,1}$. In fact, $p_{0,1,1}=0$ with the degree of $\gamma^{(0)}_{0,1}(k)$ being equal to $0$ and each iteration increases the order by $1$ as an effect of the inverse of the finite difference operator and by an additional $1$ in case the finite difference polynomial features the term $(k-l)\beta^{(0)}_{l,n}D^2w(Dw)^{-1}$, which can be detected by considering whether $p_{l+1,n,1}$ is decreased or not with respect to $p_{l,i,1}$, where $i$ is the index identifying the generator. Finally
\begin{equation}\label{eq:polyb0}
	\gamma^{(0)}_{l,n}(k)=\theta_{l,n}\prod_{m=0}^{l-p_{l,n,1}-1}(k-m)
	\nonumber
\end{equation}
So, the computation could be reduced to finding the coefficients $\theta_{l,n}$ only. The knowledge about the position of the polynomial zeros is quite useful and interesting from a theoretical point of view, nevertheless the algorithm shown in the previous subsection is more practical since it gives the polynomial coefficients rather than its zeros.
}{}

\ifthenelse{\boolean{\addtab}}{
\begin{table}
	\centering
	\caption{%
		Expressions of $\beta^{(b)}_{l,n}$ and $\gamma^{(b)}_{l,n}$ for $l=0,\ldots\,3$ and $b=1$.
	}%
	\begin{tabular}{p{.2cm}p{2cm}l}
		\toprule
		$l,n$ & $\beta^{(1)}_{l,n}$	& $\gamma^{(1)}_{l,n}(k)$\\
		\midrule
		$0,1$ & $1$					& $1$\\
		\midrule
		$1,1$ & $D^2w$				& $\frac{1}{2}(k^2+k)$\\
		\midrule
		$2,1$ & $(Dw)^{-1}(D^2w)^2$	& $\frac{1}{8}(k^4-2k^3-k^2+2k)$\\
		\midrule
		$2,2$ & $D^3w$				& $\frac{1}{6}(k^3-k)$\\
		\midrule
		$3,1$ & $(Dw)^{-2}(D^2w)^3$	& $\frac{1}{48}(k^6-9k^5+25k^4-15k^3-26k^2+24k)$\\
		\midrule
		$3,2$ & $(Dw)^{-1}D^2wD^3w$	& $\frac{1}{12}(k^5-5k^4+5k^3+5k^2-6k)$\\
		\midrule
		$3,3$ & $D^4w$				& $\frac{1}{24}(k^4-2k^3-k^2+2k)$\\
		\bottomrule
	\end{tabular}
	\label{tab:betagammab1}
\end{table}
}{}

\ifthenelse{\boolean{\addpar}}{
The case $b=1$ is slightly different. The polynomial $\gamma^{(1)}_{l,n}(k)$ has $l-p_{l,n,1}-1$ zeros placed on $0,\ldots,l-p_{l,n,1}-2$. Since the order of the polynomial is $l-p_{l,n,1}$, one zero has to be identified. To this aim, we can consider the following equivalence
\begin{equation}\label{eq:defuncb0to1}
	D^{k}\phi_a^{(1)}=a^{-1}D^{k+1}\phi_a^{(0)}.
	\nonumber
\end{equation}
It follows
\begin{multline}
	\sum_{l=0}^{k+1}\sum_{n=1}^{|\Omega_{l}|}\gamma_{l,n}^{(0)}(k+1)a^{k-l}
	(Dw)^{k+1-l+p_{l,n,1}}\prod_{m=2}^{l+1}(D^mw)^{p_{l,n,m}}
	\\
	=
	\sum_{l=0}^{k}\sum_{n=1}^{|\Omega_{l}|}\gamma_{l,n}^{(1)}(k)a^{k-l}
	(Dw)^{k+1-l+p_{l,n,1}}\prod_{m=2}^{l+1}(D^mw)^{p_{l,n,m}}.
	\nonumber
\end{multline}
Then it must be
\begin{equation}\label{poly0to1}
	\gamma_{l,n}^{(1)}(k)=\gamma_{l,n}^{(0)}(k+1)
	\qquad
	l=0,\ldots,k
	\nonumber
\end{equation}
hence $\gamma^{(1)}_{l,n}(k)$ must have $l-p_{l,n,1}$ zeros placed on $-1,\ldots,l-p_{l,n,1}-2$. It is worth noting that in the above equivalence the left hand side has to be $0$ for $l=k+1$, thus giving
\begin{equation}\label{eq:derfunczeroposb1}
	\sum_{n=1}^{|\Omega_{l}|}\beta^{(1)}_{k+1,n}\gamma_{k+1,n}^{(0)}(k+1)=0
	\quad
	\Rightarrow
	\quad
	\gamma_{k,n}^{(0)}(k)=0,k>0
	\nonumber
\end{equation}
which is always verified since we already showed that $\gamma_{k,n}^{(0)}$ has zeros placed in $0,\ldots,k-p_{k,n,1}$ and $p_{k,n,1}\leq{}0$.
}{}

\section{Performances}
\label{sec:results}

In order to prove the effectiveness of the proposed approach, the behaviour of all the operators which have been presented has to be analysed in terms of reconstruction accuracy by evaluating the norm of the error matrix. A major distinction has to be done according to the warping map regularity. If no specific constraint is posed in the design, \ac{TW} and \ac{FW} maps belong to $\mathcal{C}^0$ and $\mathcal{C}^1$ respectively as a consequence of the periodization at the boundaries, as shown in \ref{subsec:perspace}. Nevertheless, examples of a frequency map featuring a \emph{knee} or a time map having a continuous derivative value at the boundaries could be provided. For the sake of this comparison, we will simply refer to the map continuity class $\mathcal{C}^\sigma$ rather than the specific \ac{TW} or \ac{FW} continuity class. In terms of reconstruction performances, another relevant parameter is represented by the transform normalized redundancy, i.e. $M/(N\max{}Dw)$. In more detail, the analysis of the asymptotic behaviour of the reconstruction accuracy with respect to $M\to\infty$ allows to extrapolate accurate estimations for most operators, although for application purposes the number of output samples has to be kept as close as possible to $N\max{}Dw$ (see \eqref{eq:redun}). For this reason, error norms will be analysed according to a normalized redundancy interval being large enough to observe the asymptotic behaviour, namely $M/(N\max{}Dw)\in[1,10]$, whereas the interesting interval for application purposes is typically $M/(N\max{}Dw)\in[1,2]$.

\begin{figure}[t]
	\centering%
	\includegraphics[scale=.75,trim=0 10pt 0 10pt]{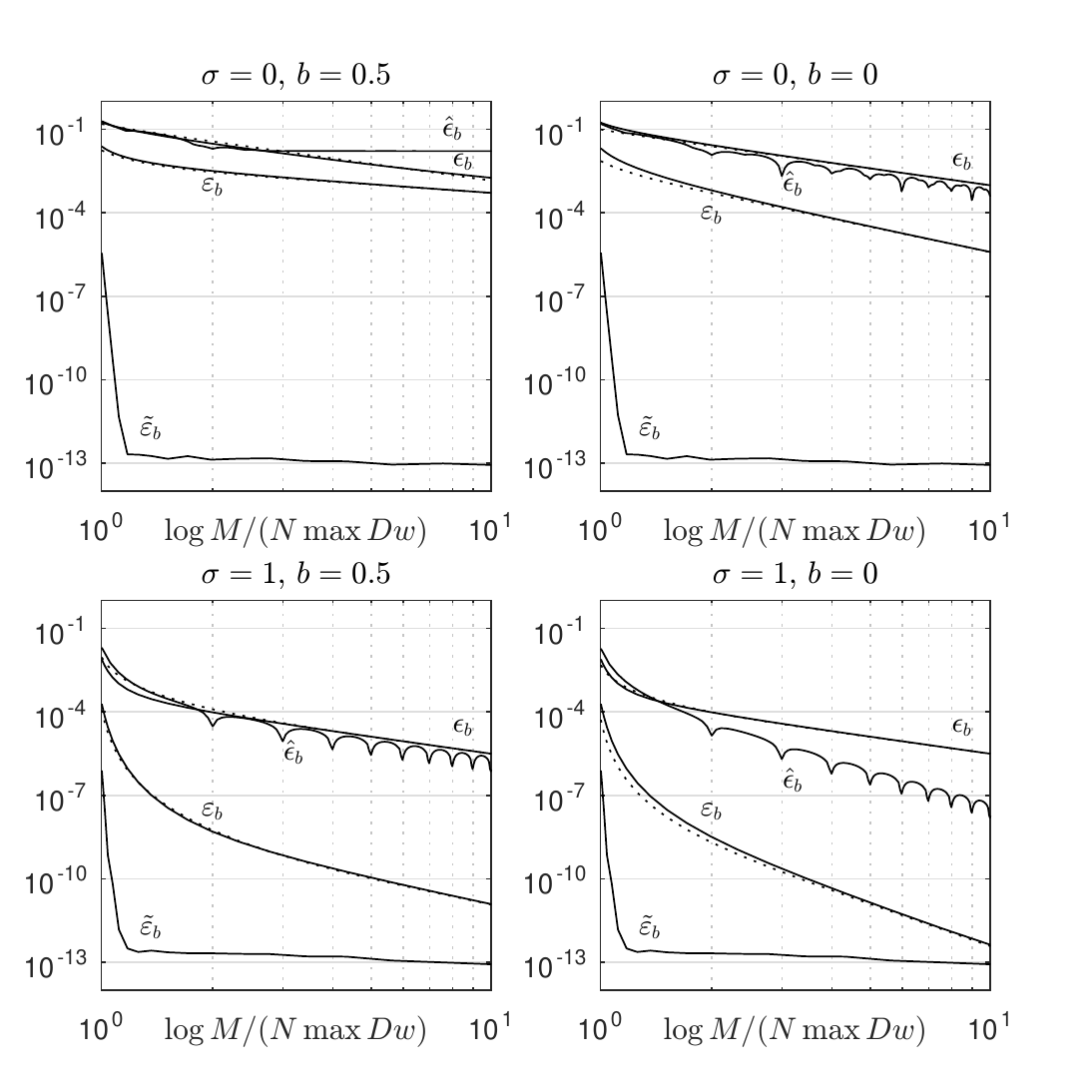}%
	\caption{%
		Reconstruction error matrix norms relative to \ac{SWF} and \ac{SAF} warping operators inverted by means of their transpose operators ($\epsilon_b$ and $\varepsilon_b$) and the operator built with the inverse map ($\hat{\epsilon}_b$) and the \ac{SAF} dual operator ($\tilde{\varepsilon}_b$). Computed values are plotted in solid lines, while estimated values for $\epsilon_b$ and $\varepsilon_b$ are plotted in dashed lines.
	}%
	\label{fig:warpperf}
\end{figure}

So, we consider the following norms
\begin{eqnarray}
	\label{eq:nrm1}
	\hat{\epsilon}^{(b)}(\sigma)
	&=&
	\|\IXX^{(b)\dag}\XX^{(b)}-\op{I}\|
	\\[3pt]
	\label{eq:nrm2}
	\epsilon^{(b)}(\sigma)
	&=&
	\|\XX^{(1-b)\dag}\XX^{(b)}-\op{I}\|
	\\[3pt]
	\label{eq:nrm3}
	\varepsilon^{(b)}(\sigma)
	&=&
	\|\WX^{(1-b)\dag}\WX^{(b)}-\op{I}\|
	\\
	\label{eq:nrm4}
	\tilde{\varepsilon}^{(b)}(\sigma)
	&=&
	\|\IWX^{(1-b)\dag}\WX^{(b)}-\op{I}\|
\end{eqnarray}
for $b$ equal to $0$ and $\nicefrac{1}{2}$ and $\sigma$ equal to $0$ and $1$, although parametric estimations with respect to $\sigma$ will be provided. In fact, thanks to the proposed models, the behaviour of \eqref{eq:nrm1}-\eqref{eq:nrm4} can be accurately foreseen. In the following, we will improve and expand the error analysis proposed in~\cite{Caporale2009}. Fig.~\ref{fig:warpperf} shows the above listed errors with respect to the normalized redundancy $M/(N\max{}Dw)$. Before going through their analytical characterization, we provide some qualitative considerations. As a first remark, it can be noticed that the errors \eqref{eq:nrm1}-\eqref{eq:nrm3} feature a linear asymptotic behaviour over a \emph{loglog} representation. 
As a consequence, understanding the relationship between the error slope and the map regularity is a major requirement for the comparison between the various warping operators. As a second remark, it can be noticed that the errors \eqref{eq:nrm1}-\eqref{eq:nrm3} exhibit a qualitative decrease by moving in Fig.~\ref{fig:warpperf} between subfigures from left to right and from top to bottom. This effect is due to the increase in regularity (top to bottom) and to the fact that the first column of matrix $\op{S}^{(b)}$ becomes null for $b\to{}0$ (left to right). These behaviours will be inspected later in the discussion for the determination of their asymptotic characterization. An additional remark concerns the periodic behaviour of $\hat{\epsilon}_b$ of period $N\max{}Dw$, which can be explained by considering that the operator $\IXX^{(b)}$, as well as $\XX^{(b)}$, features aliasing. Aliasing of $\XX^{(b)}$ happens as an effect of the periodic summation along the column index with period $M$, whereas aliasing of $\IXX^{(b)}$ is an effect of periodic summation along the row index with period $N$ (see Fig.~\ref{fig:warpsparse}), which reflects on a consistent behaviour every time $M$ is increased by multiples of $N\max{}Dw$. As a final qualitative remark, we highlight the saturation of $\hat{\epsilon}_b$ for $b=\nicefrac{1}{2}$ and $\sigma=0$ (top left subfigure in Fig.~\ref{fig:warpperf}). Again, this happens as an effect of the aliasing in $\IXX^{(b)}$. In more detail, since $N<M$, the model \eqref{eq:aliasdec} does not apply. Nevertheless, by finding a suitable integer $L$ such that $LN>M\max{}Dv$ with $v=w^{-1}$, the model \eqref{eq:aliasdec} applies with period $LN$, then the proper aliasing can be obtained by further performing a periodic summation with period $N$. When $\sigma=0$, the decay of the aliasing is such that the further periodic summation produces a constant matrix causing the constant error. This effect does not arise for $b=0$ as in \eqref{eq:nrm1} both the direct and the inverse operators feature no orthogonalization factor, which makes them both belong to $\mathcal{C}^{1}$.

As far as analytical error models are concerned, we start by modeling $\varepsilon_b$. By recalling again that $\op{I}-\op{W}_\mathrm{f}^{(\bar{b})\dag}\op{W}^{(b)}_\mathrm{f}=\op{E}^{(\bar{b})\dagger}\op{E}^{(b)}$ and exploiting \eqref{eq:tailsdec}, one gets
\begin{equation}
	\varepsilon^{(b)}(\sigma)
	\simeq
	\|\op{V}^\prime\op{S}^{(\bar{b})\dag}\op{Y}^{\prime}\op{Y}\,\op{S}^{(b)}\op{V}\|
	\label{eq:normfact}
\end{equation}
where, among the $I$ singularities contributing to the decomposition \eqref{eq:tailsdec}, the one positioned in the generic point $\xi$ having minimum regularity and maximum step has been selected. The exact asymptotic behaviour of $\varepsilon^{(b)}(\sigma)$ for $M\to\infty$ can be analytically determined by taking advantage of the knowledge of the asymptotic behaviour of $\op{S}^{(b)}$, then a very fine estimation can be heuristically obtained\appref{ (see Appendix \ref{subapp:normsaf})}:
\begin{equation}
	\varepsilon^{(b)}(\sigma)
	\simeq
	\frac{\varrho_{\varepsilon}^{(b)}(\nicefrac{M}{N})\,\Delta^{(b)}\Delta^{(\bar{b})}\varsigma^{(b)}\varsigma^{(\bar{b})}}{\pi^{2\sigma+2}(2\sigma+1+c)(1+2c)^{\nicefrac{1}{2}}}
	\frac{N^{1+c}}{M^{2\sigma+1+c}}.
	\label{eq:norm3est}
\end{equation}
where $c=2(\delta_b+\delta_{\bar{b}})$, $\Delta^{(b)}=|(Dw(\xi^+))^{2\delta_b}\beta^{(b)}_{\sigma,\Omega_\sigma}(\xi^+)-(Dw(\xi^-))^{2\delta_b}\beta^{(b)}_{\sigma,\Omega_\sigma}(\xi^-)|$ (not to be confused with the operator employed in 
\eqref{eq:coefupdateconvdiff}), $\varsigma^{(b)}=\gamma^{(b)}_{\sigma,\Omega_\sigma}(\sigma+2\delta_b)$, which, for $\sigma=0$, is equal to $1$, whereas, for $\sigma>0$, is equal to $\sigma+2$, $\nicefrac{1}{2}$ and $1$ for $b=0$, $b=\nicefrac{1}{2}$ and $b=1 $ respectively. Finally, $\varrho_{\varepsilon}^{(b)}$ is a function asymptotically equal to $1$ which takes into account the effect of the polynomial $\gamma^{(b)}_{\sigma,\Omega_\sigma}$ before it converges to its value in $\sigma+c$. $\varrho_{\varepsilon}^{(b)}$ depends only on $M/N$, $Dw(x^+)$ and $Dw(x^-)$ (or the same value $Dw(x)$ when $\sigma>0$), hence it can be preevaluated or parametrically approximated.

The estimation of $\epsilon_b(\sigma)$ is tackled in a similar way by first referring to  $\XF^{(b)}$ and $\XF^{(\bar{b})}$ and expanding both as $\WF^{(b)}+\op{A}^{(b)}$ and $\WF^{(\bar{b})}+\op{A}^{(\bar{b})}$ thus getting $\op{E}^{(\bar{b})\dagger}\op{E}^{(b)}+
\op{A}^{(\bar{b})\dag}\WF^{(b)}+
\WF^{(\bar{b})\dag}\op{A}^{(b)}+
\op{A}^{(\bar{b})\dag}\op{A}^{(b)}$
%
%
where the first and the last term are expected to be negligible, such that
\begin{equation*}
	\epsilon^{(b)}(\sigma)\simeq
	\|\op{A}^{(\bar{b})\dag}\WF^{(b)}\|+
	\|\WF^{(\bar{b})\dag}\op{A}^{(b)}\|.
\end{equation*}
In case $b=\nicefrac{1}{2}$ the two terms are equal, whereas in case $b=0$ the first term is dominant since $\op{A}^{(1)}$ refers to $\WF^{(1)}$ which features the orthogonalization factor, hence properly belongs to $\mathcal{C}^{\sigma}$, whereas $\WF^{(0)}$ practically belongs to $\mathcal{C}^{\sigma+1}$. Hence, by taking advantage of \eqref{eq:aliasdec} and introducing the multiplicity factor $\rho(b)$ equal to $2$ for $b=(0,1)$ and equal to $1$ for $b=\{0,1\}$, we get
\begin{equation}
	\epsilon^{(b)}(\sigma)\simeq
	\rho(b)\,
	\|\op{V}^\prime\op{S}^{(\bar{b})\dag}\op{U}^{\prime}\WF^{(b)}\WF^{(b)\dag}\op{U}\,\op{S}^{(\bar{b})}\op{V}
	\|^{\nicefrac{1}{2}}
	\label{eq:normfact2}
\end{equation}
whose form is very similar to \eqref{eq:normfact}. By setting $\eta=\mathrm{mod}_2(\sigma+1)$, under certain conditions, the following estimation is obtained\appref{ (see Appendix \ref{subapp:normswf})}
\begin{equation}
	\epsilon^{(b)}(\sigma)
	\simeq
	\rho(b)
	\frac{\lambda\,\vartheta^{(b)}(\varrho_\epsilon^{(b)}(\nicefrac{M}{N})\,\Delta^{(b)}\Delta^{(\bar{b})}\varsigma^{(b)}\varsigma^{(\bar{b})})^{\nicefrac{1}{2}}}{\pi^{\sigma+1}\,2^{\sigma+1+\eta}\,3^{\nicefrac{\eta}{2}}}
	\frac{N^{1+\eta}}{M^{\sigma+1+\eta}}
	\label{eq:norm2est}
\end{equation}
where $\lambda=(2\pi)^{(\sigma+1+\eta)}/(\sigma+1+\eta)!\,\mathrm{ber}(\sigma+1+\eta)$ being $\mathrm{ber}(n)$ the Bernoulli number sequence, $\vartheta^{(b)}=(((Dw(0^+))^b+(Dw(0^-))^b)/2)$ and $\varrho_\epsilon^{(b)}$ plays for $\epsilon_b(\sigma)$ the same role that $\varrho_\varepsilon^{(b)}$ plays for $\varepsilon_b(\sigma)$ in \eqref{eq:norm3est}. The coefficient $\lambda$ has been normalized so that it is nearly constant and $\lambda=2$ for $\sigma\to\infty$. For $\sigma=0,1$, $\lambda\approx3.29$, while for $\sigma=2,3$, $\lambda\approx2.16$.

Estimation \eqref{eq:norm3est} and \eqref{eq:norm2est} are also plotted in Fig.~\ref{fig:warpperf} in dotted lines and in most cases are hardly distinguishable from the corresponding computed values. For the sake of completeness, it has to be mentioned that for $\sigma=0$ and $b=1/2$ the estimation \eqref{eq:norm2est} and \eqref{eq:norm3est} decrease as $\propto{}M^{-2}$ and $\propto{}M^{-1}$ respectively, so this would be the only case where $\epsilon^{(b)}$ can be smaller that $\varepsilon^{(b)}$ for some $M$. This is actually not happening as the decomposition of  $\epsilon^{(b)}(\sigma)$ comprehends $\op{E}^{(\bar{b})\dagger}\op{E}^{(b)}$ which has been previously discarded, hence the proper estimation is represented by $\epsilon^{(b)}(\sigma)\simeq\rho(b)\|\op{A}^{(\bar{b})\dag}\WF^{(b)}\|+\varepsilon^{(b)}(\sigma)$.

As far as $\hat{\epsilon}^{(b)}$ is concerned, an exact estimation cannot be provided. Nevertheless, some practical considerations can be done. As far as the case $b=\nicefrac{1}{2}$, it is reasonable to expect that $\hat{\epsilon}^{(\nicefrac{1}{2})}$ has a behaviour similar to ${\epsilon}^{(\nicefrac{1}{2})}$, since both operators $\XX^{(\nicefrac{1}{2})}$ and $\IXX^{(\nicefrac{1}{2})}$ feature aliasing and their regularity is the same. This inference is confirmed by the measurements reported in Fig.~\ref{fig:warpperf} (apart from previously discussed case $\sigma=0$ featuring a saturation). When considering $b=0$, it can be expected that $\hat{\epsilon}^{(0)}$ is approximated by $\|\WF^{(0)\dag}\op{A}^{(0)}\|$ as $\widehat{\op{X}}_\mathrm{x}^{(0)}$ comprehends $\WX^{(0)}$ in its decomposition. This model would actually match the slope of $\hat{\epsilon}^{(0)}$ in right-bottom subfigure in the Fig.~\ref{fig:warpperf}. Nevertheless, a proper proof involving the effect of the structure of $\IXX^{(b)}$ would be needed.

We now provide some considerations about $\tilde{\varepsilon}^{(b)}$. First, we highlight that the error curve does not change with $b$. In fact, operators for $b\neq\nicefrac{1}{2}$ have been obtained by taking advantage of the approach developed for $b=\nicefrac{1}{2}$ by only redistributing the weight of the orthogonalization factor between the direct and the inverse operator. Furthermore, the error curves saturates to machine error for $M$ slightly larger than $N\max{}Dw$. Obviously, this happens as the operator $\IWX^{(1-b)}$ has been designed as pseudoinverse of operator $\WX^{(1-b)}$. The actual decrease rate $\tilde{\varepsilon}^{(b)}$ depends on the truncation operated on $\op{S}^{(b)}$ (see~\cite{Caporale2014} fore details).

Finally, \appref{we discuss some remarks about the considered operators and their performances. In particular,} we want to focus on the actual advantages carried by adopting $\WX^{(b)}$ rather than $\XX^{(b)}$ as direct operator and $\IWX^{(1-b)}$ or $\WX^{(1-b)}$ rather than $\XX^{(1-b)}$ or $\IXX^{(b)}$ as inverse operator. Both the errors $\hat{\epsilon}^{(b)}$, ${\epsilon}^{(b)}$ are quite unsatisfactory for both \ac{TW} ($\sigma=0$) and \ac{FW} ($\sigma=1$), hence the adoption of $\WX^{(b)}$ comes as a natural choice. By comparing $\epsilon^{(b)}$ and $\varepsilon^{(b)}$ for $M\gtrsim{}N\max{}Dw$, $\epsilon^{(b)}\propto{}M^{-\sigma}$ and $\varepsilon^{(b)}\propto{}M^{-2\sigma}$, thus $\varepsilon^{(b)}/\epsilon^{(b)}\simeq{}M^{-\sigma}$, meaning that the advantage of employing $\WX^{(b)}$ rather than $\XX^{(b)}$ is much more remarkable for \ac{FW} than for \ac{TW}. In fact, ${\varepsilon}^{(b)}$ might be considered sufficiently accurate for \ac{FW}, whereas for \ac{TW} not much improve in accuracy is carried out. Hence, the employment of the dual operator $\IWX^{(1-b)}$ is particularly tailored for warping and interpolation operations in the time domain.

\section{Conclusion}
\label{sec:concl}

In this work, a novel approach to time warping supporting a fast algorithm with fast inverse has been introduced by revisiting a mathematical framework previously introduced for frequency warping. Moreover, the same approach has been extended to be applicable to pure time interpolation. The method is demonstrated to be computationally effective and accurate and to carry major advantages with respect to other available techniques.

\appendices

\ifthenelse{\boolean{\addapp}}{%

\section{}
\label{app:derfunciter}

Proof of relationships \eqref{eq:derfuncinit}-\eqref{eq:derfuncdiff}. We start by setting
\begin{equation}\label{eq:derfuncfact}
	D^k\phi^{(b)}_a=\mathrm{e}^{aw}\,\psi_{a,k}^{(b)}
	\nonumber
\end{equation}
hence the following is easily verified
\begin{equation}\label{eq:derfuncfactiter}
	\psi_{a,k}^{(b)}=D\psi_{a,k-1}^{(b)}+\psi_{a,k-1}^{(b)}(aDw).
\end{equation}
Since $\psi_{a,0}^{(b)}$ is a function of degree $0$ with respect to $(aDw)$, $\psi_{a,k}^{(b)}$ must be a function of degree $k$, so we can adopt the following polynomial expansion
\begin{equation}\label{eq:derfuncfactdec}
	\psi_{a,k}^{(b)}=\sum_{l=0}^{k}\alpha_{k,k-l}^{(b)}(aDw)^l
\end{equation}
where $\alpha_{k,k-l}^{(b)}$ does not depend on $a$. By recursively applying the equivalence above, we also get
\begin{equation}\label{eq:derfuncfactdeciter}
	\psi_{a,k}^{(b)}=\psi_{a,0}^{(b)}(aDw)^k+\sum_{l=0}^{k-1}D\psi_{a,k-l-1}^{(b)}(aDw)^l
	\nonumber
\end{equation}
meaning that $\alpha_{k,0}^{(b)}=(Dw)^b$ $\forall{}k$, as stated by \eqref{eq:derfuncinit}. The following differential expansion is clearly of degree $k$ with respect to $aDw$
\begin{multline}\label{eq:derfuncfactdecdiff}
	\psi_{a,k+1}^{(b)}-\psi_{a,k}^{(b)}(aDw)=
	(\alpha_{k+1,0}^{(b)}-\alpha_{k,0}^{(b)})(aDw)^{k+1}+
	\\
	\alpha_{k+1,k+1}^{(b)}
	+
	\sum_{l=1}^{k}(\alpha_{k+1,k+1-l}^{(b)}-\alpha_{k,k+1-l}^{(b)})(aDw)^{l}.
\end{multline}
as the coefficient $\alpha_{k+1,0}^{(b)}-\alpha_{k,0}^{(b)}$ is equal to $0$ since $\alpha_{k,0}^{(b)}=(Dw)^b$. An alternative expression can be obtained by deriving \eqref{eq:derfuncfactdec} and noticing from \eqref{eq:derfuncfactiter} that $\psi_{a,k+1}^{(b)}-\psi_{a,k}^{(b)}(aDw)=D\psi_{a,k}^{(b)}$
\begin{multline}
	D\psi_{a,k}^{(b)}=D\alpha_{k,k}^{(b)}+
	\\
	\sum_{l=1}^{k}(D\alpha_{k,k-l}+l\alpha_{k,k-l}(Dw)^{-1}D^2w)(aDw)^{l}.
	\label{eq:derfuncfactdecder}
\end{multline}
then, by pairing the expansion coefficients between \eqref{eq:derfuncfactdecdiff} and \eqref{eq:derfuncfactdecder} we get equations \eqref{eq:derfunciter} and \eqref{eq:derfuncdiff}.
\section{}
\label{app:derfuncdec}

In order to validate relationships \eqref{eq:derfuncdec}-\eqref{eq:derfuncpoly} we employ the induction principle, hence we show that the validity for $k$ implies the validity for $k+1$. Moreover, since we state that $\Omega_k$ in equation \eqref{eq:derfuncdec} lists all the sequences satisfying property \eqref{eq:derfuncpow}, we have to show that there are no other functions than $\alpha_{k,l}^{(b)}$ having the same form. The proof is split in two parts. We aim to prove that (i) equation \eqref{eq:derfunciter} implies the decomposition \eqref{eq:derfuncdec} for to $\alpha_{k,k}^{(b)}$ and that (ii) equation \eqref{eq:derfuncdiff} implies the decomposition \eqref{eq:derfuncdec} for $\alpha_{k,l}^{(b)}$.

\subsection{The Functional Decomposition}
\label{subapp:derfuncdecfunc}

By deriving $\alpha_{k,k}^{(b)}$ we get
\begin{multline}
	\alpha_{k+1,k+1}^{(b)}=(Dw)^b\sum_{n=1}^{|\Omega_k|}\gamma_{k,n}^{(b)}(k)
	\sum_{l=1}^{k+1}(p_{k,n,l}+b\delta_{l-1})
	\\
	\prod_{m=1}^{k+2}(D^mw)^{p_{k,n,m}-\delta_{m-l}+\delta_{m-l-1}}.
	\nonumber
\end{multline}
In the expression above, for each couple $k,n$ an expansion in $(k+1)$ products of sequences is considered although some of them might be null because of the multiplying coefficient $p_{k,n,l}+b\delta_{l-1}$ (this is the same as saying that the derivative of $(D^mw)^0=0\,(D^{m}w)^{-1}D^{m+1}w$). Moreover, when $p_{k,n,m}=0$, for $l=m$ we get $p_{k,n,m}-\delta_{0}+\delta_{-1}=-1$, which is not compliant to the sign condition on $p_{k,n,m}$. 
Therefore, we take advantage of the definition \eqref{eq:gener} identifying non-null elements of $p_{k,n,l}+b\delta_{l-1}$. In our convention, we always assume $\Lambda_{k,n}(1)=1$, i.e. the derivative operator is always applied to the factor $Dw$ of $\alpha_{k,k}^{(b)}$ as a consequence of the presence factor $(Dw)^b$. If $b=0$, this should not always happen. Keeping $b$ as a symbolic variable and then replacing it with actual value is the \emph{trick} which is adopted here to allow a unified representation for all cases in $b\in[0,1]$. From a mathematical point of view, the case $b=0$ can be intended as $b\to{}0$. We get
\begin{multline}
	\alpha_{k+1,k+1}^{(b)}=
	\\
	(Dw)^b\sum_{n=1}^{|\Omega_k|}\gamma_{k,n}^{(b)}(k)\!\!
	\sum_{l=1}^{|\Lambda_{k,n}|}(p_{k,n,\Lambda_{k,n}(l)}+b\delta_{\Lambda_{k,n}(l)-1})
	\\
	\prod_{m=1}^{k+2}(D^mw)^{p_{k,n,m}-\delta_{m-\Lambda_{k,n}(l)}+\delta_{m-\Lambda_{k,n}(l)-1}}
	\label{eq:appexp}
\end{multline}
The new expression is now compatible with the expression for $\alpha_{k+1,k+1}^{(b)}$. In fact
\begin{multline}
	\sum_{m=1}^{k+2}(p_{k,n,m}-\delta_{m-\Lambda_{k,n}(l)}+\delta_{m-\Lambda_{k,n}(l)-1})m
	=\\
	\sum_{m=1}^{k+1}p_{k,n,m}m+\sum_{m=1}^{k+2}(-\delta_{m-\Lambda_{k,n}(l)}+\delta_{m-\Lambda_{k,n}(l)-1})\,m
	=\\
	k-l+l+1=k+1.
	\nonumber
\end{multline}
We must also show that there are no other sequences compliant to the given property which are not generated by this process. Since the generation process is invertible, if there was such a sequence, then the reverse generation process could be recursively performed till $p_{0,n,m}$. It holds $p_{0,n,1}=0$ which is satisfied by a single \emph{sequence}, i.e. $p_{0,1,1}=0$. By properly merging the two summations in \eqref{eq:appexp} by means of the expansion indexes \eqref{eq:idxexp} and taking advantage of definitions \eqref{eq:powupdate} and \eqref{eq:powupdateidxrep}, one gets the sequences $\tilde{p}_{k+1,n}$ 
\begin{multline}
	\alpha_{k+1,k+1}^{(b)}=
	(Dw)^b\sum_{n=1}^{|\Omega_k|}\gamma_{k,n}^{(b)}(k)
	\\
	\sum_{l=1}^{|\Lambda_{k,n}|}(p_{k,n,\Lambda_{k,n}(l)}+b\delta_{\Lambda_{k,n}(l)-1})
	\prod_{m=1}^{k+2}(D^mw)^{g_{k,n,l,m}}
	\\
	=(Dw)^b\sum_{n=1}^{|\tilde{\Omega}_{k+1}|}\tilde{\gamma}_{k+1,n}^{(b)}(k+1)\prod_{m=1}^{k+2}(D^mw)^{\tilde{p}_{k,n,m}}
	\nonumber
\end{multline}
where $\tilde{\gamma}_{k+1,n}^{(b)}(k+1)$ is a suitable coefficient derived by applying the expansion \eqref{eq:idxexp}. The set $\tilde{p}_{k+1}$ can feature repeated sequences. This can be proved by applying the inverse generative operation and showing that a sequence ${p}_{k+1,n}$ can originate from different sequence from the set ${p}_{k}$. In fact, the inverse generative operator $\delta_{m-l}-\delta_{m-l-1}$ can be applied to all indexes $l$ such that $p_{k+1,n,l+1}>0$, hence all sequences featuring at least two positive elements have more than one generator. By finally applying the index contraction \eqref{eq:powupdateidxrep} we get an expression for $\alpha_{k+1,k+1}^{(b)}$ which is compliant to \eqref{eq:derfuncdec} with respect to the functional part expressed by \eqref{eq:derfuncpow} and \eqref{eq:derfuncprod}. The coefficients $\gamma_{k+1,n}^{(b)}(k+1)$ could be also found from $\tilde{\gamma}_{k+1,n}^{(b)}(k+1)$ by exploiting \eqref{eq:derfuncdiff}. However, this is not useful as a more general problem is solved in Appendix \ref{subapp:derfuncdecpoly} by finding $\alpha_{k,l}^{(b)}$ rather than its samplings for $l=k$, that is finding $\gamma_{l,n}^{(b)}(k)$ rather than $\gamma_{k,n}^{(b)}(k)$.

\subsection{The Polynomial Decomposition}
\label{subapp:derfuncdecpoly}

Equation \eqref{eq:derfuncdec} sets the decoupling between $x$ and the derivative order. We must show that it holds according to the \eqref{eq:derfuncdiff}. Functions $\gamma_{l,n}^{(b)}(k)$ are expected to be polynomials in $k$. The relationship \eqref{eq:derfuncdiff} does not involve any transformation with respect to the variable $x$, hence the functional part in $x$ has to be the same on the left and on the right hand side. Considering that $\alpha_{k,k}^{(b)}$ already implies $\alpha_{k+1,k+1}^{(b)}$ and that $\alpha_{k+1,0}^{(b)}=\alpha_{k,0}^{(b)}$, showing that the validity of $\alpha_{k,l}^{(b)}$ and $\alpha_{k,l+1}^{(b)}$ implies the validity of $\alpha_{k+1,l+1}^{(b)}$ is sufficient to prove that the decomposition \eqref{eq:derfuncdec} holds for every $\alpha_{k,l}^{(b)}$ by the induction principle.

By deriving \eqref{eq:derfuncdec} and exploiting the proof in \ref{subapp:derfuncdecfunc} we get
\begin{multline}
	D\alpha_{k,l}^{(b)}=(Dw)^b\sum_{n=1}^{|\Omega_l|}\gamma_{l,n}^{(b)}(k)
	\sum_{q=1}^{|\Lambda_{l,n}|}(p_{l,n,\Lambda_{l,n}(q)}+b\delta_{\Lambda_{l,n}(q)-1})
	\\
	\prod_{m=1}^{l+2}(D^mw)^{p_{l,n,m}-\delta_{m-\Lambda_{l,n}(q)}+\delta_{m-\Lambda_{l,n}(q)-1}}
	\label{eq:appexp2}
\end{multline}
which is the same as \eqref{eq:appexp} apart from replacing $k$ with $l$ in the second index with. For $\alpha_{k,l}^{(b)}D^2w(Dw)^{-1}$ we get
\begin{multline}
	\alpha_{k,l}^{(b)}D^2w(Dw)^{-1}=
	\\
	(Dw)^b\sum_{n=1}^{|\Omega_l|}\gamma_{l,n}^{(b)}(k)\prod_{m=1}^{l+1}
	(D^mw)^{p_{l,n,m}-\delta_{m-1}+\delta_{m-2}}
	\nonumber
\end{multline}
which can be expanded to fit the representation of \eqref{eq:appexp2}
\begin{multline}
	\Delta_k\alpha_{k+1,l+1}^{(b)}=
	(Dw)^b\sum_{n=1}^{|\Omega_l|}\gamma_{l,n}^{(b)}(k)
	\\
	\sum_{q=1}^{|\Lambda_{l,n}|}(p_{l,n,\Lambda_{l,n}(q)}+(b+k-l)\delta_{\Lambda_{l,n}(q)-1})
	\\
	\prod_{m=1}^{l+2}(D^mw)^{p_{l,n,m}-\delta_{m-\Lambda_{l,n}(q)}+\delta_{m-\Lambda_{l,n}(q)-1}}
	\nonumber
\end{multline}
where $\Delta_k$ represents the finite difference operator applied to $\alpha_{k+1,l+1}$ with respect to $k$. By taking advantage of definitions \eqref{eq:powupdate}-\eqref{eq:powupdateidx} we get
\begin{multline}
	\Delta_k\alpha_{k+1,l+1}^{(b)}=(Dw)^b\\
	\sum_{n=1}^{|\tilde{\Omega}_{l+1}|}\gamma^{(b)}_{l,\Psi_l(n)}(k)\,r_{l,\Psi_l(n),\Phi_l(n)}(k)
	\prod_{m=1}^{l+2}(D^mw)^{\tilde{p}_{l+1,n,m}}
	\nonumber
\end{multline}
which, according to \eqref{eq:coefupdateconv} can be written as
\begin{equation}
	\Delta_k\alpha_{k+1,l+1}^{(b)}=
	\sum_{n=1}^{|\tilde{\Omega}_{l+1}|}\tilde{\beta}_{l+1,n}^{(b)}\,\tilde{\gamma}_{l+1,n}^{(b)}(k)
	\nonumber
\end{equation}
with obvious definition of $\tilde{\beta}^{(b)}_{l+1,n}$. Terms of the summation sharing the same factor $\beta^{(b)}_{l+1,n}$ must be collected according to the partitioning \eqref{eq:idxpart}, hence we have
\begin{eqnarray}
	\Delta_k\alpha_{k+1,l+1}^{(b)}
	&=&
	\sum_{n=1}^{|{\Omega}_{l+1}|}\sum_{q=1}^{|{\Upsilon}_{l,n}|}
	\tilde{\beta}^{(b)}_{l+1,\Upsilon_{l,n}(q)}\tilde{\gamma}^{(b)}_{l+1,\Upsilon_{l,n}(q)}(k)
	\nonumber
	\\
	&=&
	\sum_{n=1}^{|{\Omega}_{l+1}|}{\beta}^{(b)}_{l+1,n}
	\sum_{q=1}^{|{\Upsilon}_{l,n}|}\tilde{\gamma}^{(b)}_{l+1,\Upsilon_{l,n}(q)}(k)
	\nonumber
\end{eqnarray}
where we took advantage of \eqref{eq:pownext} for moving $\tilde{\beta}^{(b)}_{l+1}$ from the second summation . If $\Delta_k\alpha_{k+1,l+1}^{(b)}$ can be represented with this decomposition with respect to $x$, then $\alpha_{k+1,l+1}^{(b)}$ must support the same decomposition as $\Delta_k$ does not affect the function in $x$. By considering the definition \eqref{eq:coefupdateconvdiff} it can be inferred that if \eqref{eq:derfuncpoly} holds for $l=0$, then every $\gamma_{l,n}$ is a polynomial in $k$. Then we get
\begin{equation}
	\Delta_k\alpha_{k+1,l+1}^{(b)}=
	\sum_{n=1}^{|{\Omega}_{l+1}|}
	\beta^{(b)}_{l+1,n}
	\Delta_k\gamma^{(b)}_{l+1,n}(k).
	\nonumber
\end{equation}
By applying the inverse of the finite difference operator, the polynomial $\Delta_k\gamma^{(b)}_{l+1,n}(k)$ remains a polynomial, hence we can confirm that $\alpha_{k+1,l+1}^{(b)}$ supports the decomposition \eqref{eq:derfuncdec} given that $\alpha_{k,l}^{(b)}$ supports it:
\begin{equation}
	\alpha_{k+1,l+1}^{(b)}=
	\sum_{n=1}^{|{\Omega}_{l+1}|}
	\beta^{(b)}_{l+1,n}
	\Gamma\Delta_k\gamma^{(b)}_{l+1,n}(k).
	\nonumber
\end{equation}
\section{}
\label{app:normest}

For the comparison of \eqref{eq:nrm1}-\eqref{eq:nrm4} analytical expressions are desirable. The determination of both \eqref{eq:nrm2} and \eqref{eq:nrm3}, representing the error norms for the \ac{SWF} and \ac{SAF} case respectively, is approached by taking advantage of the asymptotic behaviour of $\op{S}^{(b)}$ and $\op{S}^{(\bar{b})}$ for $M\to\infty$.

\subsection{Error norm for \ac{SAF}}
\label{subapp:normsaf}

The norm \eqref{eq:normfact} can be estimated by recalling that an accurate estimation of matrix $\op{Y}^{\prime}\op{Y}$ can be computed (see~\cite{Caporale2010b}) and that, for sufficiently large $M/(N\max{}Dw)$, $\op{S}^{(b)}$ tends to be approximated by its $\sigma$-th lower diagonal and, for $M\to\infty$, by its maxima. 

By defining $\mathrm{diag}(\cdot,n)$ as the operator that isolates the $n$-th lower diagonal and by the subscript the shift applied on the resulting vector, matrix $\op{Y}^{\prime}\op{Y}$ is such that $\mathrm{diag}(\op{Y}^{\prime}\op{Y},2n)=\op{y}_n$ where $\op{y}_0(k)=\nicefrac{M}{2k+1}$, whereas $\mathrm{diag}(\op{Y}^{\prime}\op{Y},2n+1)\approx{}0$. By setting $\op{s}^{(b)}=\mathrm{diag}(\op{S}^{(b)},\sigma)$, we get
\begin{equation*}
	\mathrm{diag}(\op{S}^{(\bar{b})\dag}\op{Y}^{\prime}\op{Y}\,\op{S}^{(b)},2n)
	\simeq
	\op{s}^{(\bar{b})}_{\max{}(0,-2n)}\cdot\op{y}_{|n|+\sigma}\cdot\op{s}^{(b)}_{\max{}(0,2n)}
\end{equation*}
while even lower and upper diagonals are null. For determining the asymptotic behaviour, by exploiting of \eqref{eq:kernelS}-\eqref{eq:kernelJ}, we get
\begin{equation*}
	\op{s}^{(b)}(k)=
	\frac{\gamma^{(b)}_{\sigma,\Omega_{\sigma}}(\sigma+k)}{M^{\sigma+1}}\left(
	\frac{\beta^{(b)}_{\sigma,\Omega_{\sigma}}(\xi^+)}{J^k(\xi^+)}-
	\frac{\beta^{(b)}_{\sigma,\Omega_{\sigma}}(\xi^-)}{J^k(\xi^-)}
	\right)
\end{equation*}
meaning that, for $M\to\infty$, $\op{s}^{({b})}_{\max{}(0,2n)}$ and $\op{s}^{(\bar{b})}_{\max{}(0,2n)}$ are fully represented by their first non-zero item. We also recall that, for $b=0$ $\op{s}^{(b)}(0)=0$, hence the first non-zero item of $\op{s}^{(b)}$ of is $\op{s}^{(b)}(c/2)$. It follows
\begin{equation*}
	\max(\mathrm{diag}(\op{S}^{(\bar{b})\dag}\op{Y}^{\prime}\op{Y}\,\op{S}^{(b)},2n))
	\propto
	J^{-(2|n|+cH(n))}
\end{equation*}
where $H$ is the Heaviside step function. Hence, for $b\in{}(0,1)$, matrix $\op{S}^{(\bar{b})\dag}\op{Y}^{\prime}\op{Y}\,\op{S}^{(b)}$ has a single maximum obtained by $n=0$, i.e. the first element of the main diagonal, i.e. $\op{s}^{(b)}(0)\,\op{s}^{(\bar{b})}(0)\,\op{y}_\sigma(0)$. Conversely, for $b=0$, the diagonals identified by $n=0$ and $n=-1$ feature maxima sharing the same order of magnitude being $\op{s}^{(0)}(1)\,\op{s}^{(1)}(1)\,\op{y}(\sigma+1)$ and $\op{s}^{(0)}(2)\,\op{s}^{(1)}(0)\,\op{y}(\sigma+1)$ respectively being also equal since $\gamma^{(0)}_{\sigma,\Omega_{\sigma}}(\sigma+1)\gamma^{(1)}_{\sigma,\Omega_{\sigma}}(\sigma+1)=\gamma^{(0)}_{\sigma,\Omega_{\sigma}}(\sigma+2)\gamma^{(1)}_{\sigma,\Omega_{\sigma}}(\sigma)$ being equal to $1$ and $\sigma+2$ for $sigma=0$ and $\sigma>0$ respectively. Finally, for $b\in(0,1)$, we obtain the following asymptotic expression
\begin{equation*}
	\op{V}^\prime\op{S}^{(\bar{b})\dag}\op{Y}^{\prime}\op{Y}\,\op{S}^{(b)}\op{V}\to
	\op{V}_0^{\prime}\op{V}_0^{\phantom{\prime}}\,\op{s}^{(\bar{b})}(0)\,\op{y}_\sigma(0)\,\op{s}^{(b)}(0)
\end{equation*}
where $\op{V}_n$ is the $n$-th row of $\op{V}$. Hence, the norm can be computed by calculating $\|\op{V}_0^{\prime}\op{V}_0^{\phantom{\prime}}\|=N$. As far as the case $b=0$ is concerned, it holds
\begin{eqnarray*}
	\op{V}^\prime\op{S}^{(1)\dag}\op{Y}^{\prime}\op{Y}\,\op{S}^{(0)}\op{V}
	&\to&
	\op{V}_0^{\prime}\op{V}_c^{\phantom{\prime}}
	\,\op{s}^{(1)}(0)\,\op{y}_\sigma(\nicefrac{c}{2})\,\op{s}^{(0)}(c)+
	\\
	&&
	\op{V}_{\nicefrac{c}{2}}^{\prime}\op{V}_{\nicefrac{c}{2}}^{\phantom{\prime}}
	\,\op{s}^{(1)}(\nicefrac{c}{2})\,\op{y}_\sigma(\nicefrac{c}{2})\,\op{s}^{(0)}(\nicefrac{c}{2})
\end{eqnarray*}
where the coefficient $c$ has been highlighted on purpose. Since $\op{V}_n$ and $\op{V}_{2n+1}$ are approximately orthogonal to each other, then $\|\op{V}_0^{\prime}\op{V}_c^{\phantom{\prime}}+\op{V}_{\nicefrac{c}{2}}^{\prime}\op{V}_{\nicefrac{c}{2}}^{\phantom{\prime}}\|=\max(\|\op{V}_0^{\prime}\op{V}_c^{\phantom{\prime}}\|,\|\op{V}_{\nicefrac{c}{2}}^{\prime}\op{V}_{\nicefrac{c}{2}}^{\phantom{\prime}}\|)\approx{}N\max(\nicefrac{1}{c+1},\nicefrac{1}{\sqrt{2c+1}})=\nicefrac{1}{\sqrt{2c+1}}$. Finally, it holds
\begin{equation*}
	\|\op{V}^\prime\op{S}^{(\bar{b})\dag}\op{Y}^{\prime}\op{Y}\,\op{S}^{(b)}\op{V}\|\to
	\frac{N}{(1+2c)^{\nicefrac{1}{2}}}
	\,\op{s}^{(\bar{b})}(\nicefrac{c}{2})\,\op{y}_\sigma(\nicefrac{c}{2})\,\op{s}^{(b)}(\nicefrac{c}{2}).
\end{equation*}
which can be used to obtain the asymptotic value in \eqref{eq:norm3est}. Finally $\varepsilon_b(\sigma)$ is heuristically estimated by replacing the maximum value of vector $\op{s}_{\nicefrac{c}{2}}^{(\bar{b})}\cdot\op{y}_{\sigma+{\nicefrac{c}{2}}}\cdot\op{s}_{\nicefrac{c}{2}}^{(b)}$ with its norm:
\begin{equation*}
	\varepsilon_b(\sigma)
	\simeq
	\frac{N}{(1+2c)^{\nicefrac{1}{2}}}\,
	\|\,
	\op{s}_{\nicefrac{c}{2}}^{(\bar{b})}\cdot\op{y}_{\sigma+{\nicefrac{c}{2}}}\cdot\op{s}_{\nicefrac{c}{2}}^{(b)}
	\|.
\end{equation*}
which is the same as \eqref{eq:norm3est}.

\subsection{Error norm for \ac{SWF}}
\label{subapp:normswf}

In order to determine an analytical expression for the asymptotic value of \eqref{eq:normfact2}, we take advantage of the similarities with \eqref{eq:normfact}. To do this, the kernel $\op{U}^{\prime}\WF^{(b)}\WF^{(b)\dag}\op{U}$ has to be estimated. From a qualitative point of view, for large $M$, $\WF^{(b)}\WF^{(b)\dag}$ is a $M\times{}M$ matrix being equal to $0$ apart from its central $N\max{}Dw\times{}N\max{}Dw$ square. Moreover, when $b=\nicefrac{1}{2}$, the central $N\mathrm{min}Dw\times{}N\mathrm{min}Dw$ square is approximately equal to the identity matrix. For $b\neq{}0$ the same square is also approximately diagonal. As fas ar $\op{U}$ si concerned, for large $M$ the central $N$ items of even and odd columns tends to be odd linear or even approximately constant vectors respectively. As a consequence, only the items of $\op{U}^{\prime}\WF^{(b)}\WF^{(b)\dag}\op{U}$ corresponding to odd row and column indexes will be remarkably larger than $0$. From a quantitative point of view we first characterize the effect of $\WF^{(b)}$ and afterwards the effect of $\op{U}$.

By assuming that the involved items of $U$ are constant, the effect of $\WF^{(b)}\WF^{(b)\dag}$ can be replaced by the sum of its items. By taking advantage of \eqref{eq:warpinf}, one gets $\sum_{m_1\in\mathds{Z}}$
\begin{equation*}
	\sum_{m_1\in\mathds{Z}_M}\sum_{m_2\in\mathds{Z}_M}\sum_{n\in\mathds{Z}_N}\op{W}(m_1,n)\op{W}^*(m_2,n)=N(Dw(0))^{2b}
\end{equation*}
where, in case $\sigma=0$ and $\xi=0$, $(Dw(0))^{2b}$ has to be intended as the square of mean value of $Dw(0^+))^b$ and $Dw(0^-))^b$, $(((Dw(0^+))^b+(Dw(0^-))^b)/2)^2=(\vartheta^{(b)})^2$. Hence, $\WF^{(b)}\WF^{(b)\dag}$ can be replaced by a windowing function asymptotically isolating the row vector $\op{U}(0,\cdot)$ and a constant, thus
\begin{equation*}
	\op{U}^{\prime}\WF^{(b)}\WF^{(b)\dag}\op{U}
	\to
	(\vartheta^{(b)})^2\,\op{U}(0,\cdot)'\op{U}(0,\cdot).
\end{equation*}

As far as the effect of $\op{U}$ is concerned, by taking advantage of \eqref{eq:mcot}, the items of $\op{U}(0,\cdot)'\op{U}(0,\cdot)$ can be obtained by properly scaling the coefficients of the Taylor series expansion of $\zeta(z)=\pi\mathrm{cot}(\pi{}z)-\nicefrac{1}{z}$, from which the Bernoulli numbers in the coefficient $\lambda$ in \eqref{eq:norm2est} originate. Given that the even items of this sequence are equal to $0$ and the odd items represent a decreasing series, it holds
\begin{equation*}
	\max(\op{S}^{(\bar{b})\dag}\op{U}^{\prime}\WF^{(b)}\WF^{(b)\dag}\op{U}\,\op{S}^{(\bar{b})})
	\to
	(\vartheta^{(b)})^2
	\op{s}^{(\bar{b})}(\eta)\,\op{u}_\sigma(\eta)\,\op{s}^{(\bar{b})}(\eta)
\end{equation*}
where $\eta=\mathrm{mod}_2(\sigma+1)$ and $\op{u}=\mathrm{diag}(\op{U}(0,\cdot)'\op{U}(0,\cdot))$. The alternating behaviour between $\op{s}^{(\bar{b})}(1)$ and $\op{s}^{(\bar{b})}(0)$ comes from the fact that even items of $\op{u}$ are equal to $0$, whereas the fact that the most significant item of $\op{U}(0,\cdot)'\op{U}(0,\cdot)$ belongs to the diagonal comes from the fact that $\op{s}^{(\bar{b})}$ nullifies the first $\sigma-1$ columns and rows of $\op{U}(0,\cdot)'\op{U}(0,\cdot)$. Going forward we get
\begin{multline*}
	\op{V}^\prime\op{S}^{(\bar{b})\dag}\op{U}^{\prime}\WF^{(b)}\WF^{(b)\dag}\op{U}\,\op{S}^{(\bar{b})}\op{V}\to
	\\
	\op{V}_\eta^{\prime}\op{V}_\eta^{\phantom{\prime}}\,
	(\vartheta^{(b)})^2
	\op{s}^{(\bar{b})}(\eta)\,\op{u}_\sigma(\eta)\,\op{s}^{(\bar{b})}(\eta)
\end{multline*}
whose norm can be obtained by considering that $\|\op{V}_\eta^{\prime}\op{V}_\eta^{\phantom{\prime}}\|=N/3^{\eta}$. Finally, by applying the same heuristic principle adopted in Appendix \ref{subapp:normsaf}, it results
\begin{equation*}
	\epsilon_b(\sigma)
	\simeq
	\rho(b)\,
	\frac{N^{\nicefrac{1}{2}}}{3^{\nicefrac{\eta}{2}}}\,
	\vartheta^{(b)}
	\|\,
	\op{s}_{\eta}^{(\bar{b})}\cdot\op{u}_{\sigma+\eta}\cdot\op{s}_{\eta}^{(\bar{b})}
	\|^{\nicefrac{1}{2}}.
\end{equation*}
which is the same as \eqref{eq:norm2est}.

}{}



\ifthenelse{\boolean{\supmat}}{%
	
\newpage
\setcounter{page}{1}

\markboth%
{Time Warping and Interpolation Operators for Piecewise Smooth Maps: \MakeLowercase{\textit{Supplemental Material}}}
{Time Warping and Interpolation Operators for Piecewise Smooth Maps: \MakeLowercase{\textit{Supplemental Material}}}
	
\section{}

\subsection{Example of polynomial computation }

\begin{table}
	\centering
	\caption{%
		Expressions of $\beta^{(b)}_{l,n}$ and $\gamma^{(b)}_{l,n}$ for $l=0,\ldots\,3$ and $b=0$.
	}%
	\begin{tabular}{p{.2cm}p{2cm}l}
		\toprule
		$l,n$ & $\beta^{(0)}_{l,n}$	& $\gamma^{(0)}_{l,n}(k)$\\
		\midrule
		$0,1$ & $1$					& $1$\\
		\midrule
		$1,1$ & $(Dw)^{-1}D^2w$		& $\frac{1}{2}(k^2-k)$\\
		\midrule
		$2,1$ & $(Dw)^{-2}(D^2w)^2$	& $\frac{1}{8}(k^4-6k^3+11k^2-6k)$\\
		\midrule
		$2,2$ & $(Dw)^{-1}D^3w$ 	& $\frac{1}{6}(k^3-3k^2+2k)$\\
		\midrule
		$3,1$ & $(Dw)^{-3}(D^2w)^3$	& $\frac{1}{48}(k^6\!-\!15k^5\!+\!85k^4\!-\!225k^3\!+\!274k^2\!-\!120k)$\\
		\midrule
		$3,2$ & $(Dw)^{-2}D^2wD^3w$	& $\frac{1}{12}(k^5-10k^4+35k^3-50k^2+24k)$\\
		\midrule
		$3,3$ & $(Dw)^{-1}D^4w$		& $\frac{1}{24}(k^4-6k^3+11k^2-6k)$\\
		\bottomrule
	\end{tabular}
	\label{tab:betagammab0}
\end{table}

\begin{table}
	\centering
	\caption{%
		Expressions of $\beta^{(b)}_{l,n}$ and $\gamma^{(b)}_{l,n}$ for $l=0,\ldots\,3$ and $b=1$.
	}%
	\begin{tabular}{p{.2cm}p{2cm}l}
		\toprule
		$l,n$ & $\beta^{(1)}_{l,n}$	& $\gamma^{(1)}_{l,n}(k)$\\
		\midrule
		$0,1$ & $1$					& $1$\\
		\midrule
		$1,1$ & $D^2w$				& $\frac{1}{2}(k^2+k)$\\
		\midrule
		$2,1$ & $(Dw)^{-1}(D^2w)^2$	& $\frac{1}{8}(k^4-2k^3-k^2+2k)$\\
		\midrule
		$2,2$ & $D^3w$				& $\frac{1}{6}(k^3-k)$\\
		\midrule
		$3,1$ & $(Dw)^{-2}(D^2w)^3$	& $\frac{1}{48}(k^6-9k^5+25k^4-15k^3-26k^2+24k)$\\
		\midrule
		$3,2$ & $(Dw)^{-1}D^2wD^3w$	& $\frac{1}{12}(k^5-5k^4+5k^3+5k^2-6k)$\\
		\midrule
		$3,3$ & $D^4w$				& $\frac{1}{24}(k^4-2k^3-k^2+2k)$\\
		\bottomrule
	\end{tabular}
	\label{tab:betagammab1}
\end{table}

\subsection{Example of computation of $\beta_{l,n}^{(b)}$}

As instance, $p_{0}=[0]$ with $\Lambda_{0,1}=\{1\}$, $p_{1}=[-1\quad{}1]^T$ with $\Lambda_{1,1}=\{1\quad{}2\}$, then
\begin{equation}
	p_{2}=
	\begin{bmatrix}
	p_{1,1} & p_{1,1}\\
	0 & 0
	\end{bmatrix}
	+
	\begin{bmatrix}
	-1 & \phantom{-}0\phantom{-}\\
	\phantom{-}1 &           -1\phantom{-}\\
	\phantom{-}0 & \phantom{-}1\phantom{-}
	\end{bmatrix}
	=
	\begin{bmatrix}
	-2 &           -1\phantom{-}\\
	\phantom{-}2 & \phantom{-}0\phantom{-}\\
	\phantom{-}0 & \phantom{-}1\phantom{-}
	\end{bmatrix}
	\nonumber
	\end{equation}
	with $\Lambda_{2,1}=\{1\quad{}2\}$ and $\Lambda_{1,2}=\{1\quad{}3\}$, then
	\begin{equation}
	\tilde{p}_{3}=
	\begin{bmatrix}
	p_{2,1} & p_{2,1} & p_{2,2} & p_{2,2} \\
	0 & 0 & 0 & 0\\
	\end{bmatrix}
	+
	\begin{bmatrix}
	-1 & \phantom{-}0 &           -1 & \phantom{-}0\phantom{-}\\
	\phantom{-}1 &           -1 & \phantom{-}1 & \phantom{-}0\phantom{-}\\
	\phantom{-}0 & \phantom{-}1 & \phantom{-}0 &           -1\phantom{-}\\
	\phantom{-}0 & \phantom{-}0 & \phantom{-}0 & \phantom{-}1\phantom{-}
	\end{bmatrix}
	\nonumber
\end{equation}
resulting in 
\begin{equation}
	p_{3}=
	\begin{bmatrix}
	-3 &           -2 &            -1\phantom{-}\\
	\phantom{-}3 & \phantom{-}1 &  \phantom{-}0\phantom{-}\\
	\phantom{-}0 & \phantom{-}1 &  \phantom{-}0\phantom{-}\\
	\phantom{-}0 & \phantom{-}0 &  \phantom{-}1\phantom{-}
	\end{bmatrix}
	\nonumber
\end{equation}
with $\Xi_2=[1\quad{}2\quad{}2\quad{}3]$. \appref{An accurate proof of this procedure supported by theoretical considerations on the uniqueness of the representation \eqref{eq:derfuncpow} is given in Appendix \ref{subapp:derfuncdecfunc}.}

\subsection{Example of computation of $\gamma_{l,n}^{(b)}$}

As instance, $u_{0,1,1}=[b\quad{}1]$, steps $2-5$ can be skipped for $l=1$, hence $c_{1,1}=[0\quad{}b-\nicefrac{1}{2}\quad{}\nicefrac{1}{2}]$. Going forward
\begin{equation}
	u_{1,\Psi_1(n),\Phi_1(n)}
	=
	\begin{bmatrix}
	b-2 && 1 \\
	1 && 0
	\end{bmatrix}
	\quad
	c_{1,\Psi_1(n)}
	=
	\begin{bmatrix}
	c_{1,1} \\
	c_{1,1}
	\end{bmatrix}
	\nonumber
\end{equation}
the convolution gives
\begin{equation}
	\tilde{u}_{2,n}
	=
	\begin{bmatrix}
	0 && b^2 -\nicefrac{5}{2}b + 1 && \nicefrac{3}{2}b-\nicefrac{3}{2} && \nicefrac{1}{2}\\
	0 && b-\nicefrac{1}{2} && \nicefrac{1}{2} && 0
	\end{bmatrix}
	\nonumber
\end{equation}
while step $5$ can be skipped because $\Xi_1$ does not have any repeated elements, hence $\tilde{c}_{2,n}=\tilde{u}_{2,n}$ and finally operator $\Gamma$ is applied thus obtaining $c_{2,n}$
\begin{equation}
	c_{2,n}
	=
	\begin{bmatrix}
	0 & -\nicefrac{b^2}{2} + \nicefrac{3}{2}b - \nicefrac{3}{4} & \nicefrac{b^2}{2} - 2b + \nicefrac{11}{8} & \nicefrac{1}{2}b - \nicefrac{3}{4} & \nicefrac{1}{8} \\
	0 & -\nicefrac{1}{2}b+\nicefrac{1}{3} & \nicefrac{1}{2}b-\nicefrac{1}{2} & \nicefrac{1}{2} & 0
	\end{bmatrix}
	\nonumber
\end{equation}
In the above example, the index $n$ refers to the row while the column index is implied. \appref{The proof supporting this procedure is given in \ref{subapp:derfuncdecpoly}.}
	
Functions $\beta^{(0)}_{l,n}$ and $\beta^{(1)}_{l,n}$ and the corresponding polynomials $\gamma^{(0)}_{l,n}$ and $\gamma^{(1)}_{l,n}$ for some values of $l$ and $n$ are given in Tables \ref{tab:betagammab0} and \ref{tab:betagammab1} respectively.

The case $b=1$ is slightly different. The polynomial $\gamma^{(1)}_{l,n}(k)$ has $l-p_{l,n,1}-1$ zeros placed on $0,\ldots,l-p_{l,n,1}-2$. Since the order of the polynomial is $l-p_{l,n,1}$, one zero has to be identified. To this aim, we can consider the following equivalence
\begin{equation}\label{eq:defuncb0to1}
	D^{k}\phi_a^{(1)}=a^{-1}D^{k+1}\phi_a^{(0)}.
	\nonumber
\end{equation}
It follows
\begin{multline}
	\sum_{l=0}^{k+1}\sum_{n=1}^{|\Omega_{l}|}\gamma_{l,n}^{(0)}(k+1)a^{k-l}
	(Dw)^{k+1-l+p_{l,n,1}}\prod_{m=2}^{l+1}(D^mw)^{p_{l,n,m}}
	\\
	=
	\sum_{l=0}^{k}\sum_{n=1}^{|\Omega_{l}|}\gamma_{l,n}^{(1)}(k)a^{k-l}
	(Dw)^{k+1-l+p_{l,n,1}}\prod_{m=2}^{l+1}(D^mw)^{p_{l,n,m}}.
	\nonumber
\end{multline}
Then it must be
\begin{equation}\label{poly0to1}
	\gamma_{l,n}^{(1)}(k)=\gamma_{l,n}^{(0)}(k+1)
	\qquad
	l=0,\ldots,k
	\nonumber
\end{equation}
hence $\gamma^{(1)}_{l,n}(k)$ must have $l-p_{l,n,1}$ zeros placed on $-1,\ldots,l-p_{l,n,1}-2$. It is worth noting that in the above equivalence the left hand side has to be $0$ for $l=k+1$, thus giving
\begin{equation}\label{eq:derfunczeroposb1}
	\sum_{n=1}^{|\Omega_{l}|}\beta^{(1)}_{k+1,n}\gamma_{k+1,n}^{(0)}(k+1)=0
	\quad
	\Rightarrow
	\quad
	\gamma_{k,n}^{(0)}(k)=0,k>0
	\nonumber
\end{equation}
which is always verified since we already showed that $\gamma_{k,n}^{(0)}$ has zeros placed in $0,\ldots,k-p_{k,n,1}$ and $p_{k,n,1}\leq{}0$.

}{}

\end{document}